\documentclass[12pt,notitlepage]{article}
\pdfoutput=1

\usepackage{amsmath}        
\usepackage{amssymb}         
\usepackage{amsfonts}        
\usepackage{graphicx}% Include figure files
\usepackage{bm}% bold math
\usepackage{xcolor}
\usepackage[normalem]{ulem}
\usepackage{booktabs}
\usepackage{multirow}
\usepackage[export]{adjustbox}
\usepackage{floatrow}
\newfloatcommand{capbtabbox}{table}[][3.5in]
\newfloatcommand{capbfigbox}{figure}[][2.3in]
\usepackage{blindtext}
\usepackage{booktabs}
\usepackage{hepunits}
\usepackage[normalem]{ulem}
\usepackage{bbm}
\usepackage{slashed}       % \slashed{k}
\usepackage[margin=2cm]{geometry}   % reasonable margins
\usepackage{cite}          % grouping citations (incompatible with collref)
\usepackage{tikz}

\usepackage[
	colorlinks=true,
	citecolor=black,
	linkcolor=black,
	urlcolor=blue,
	hypertexnames=false]{hyperref}

\newenvironment{institutions}[1][2em]{\begin{list}{}{\setlength\leftmargin{#1}\setlength\rightmargin{#1}}\item[]}{\end{list}}
\newcommand{\email}[1]{\href{mailto:#1}{#1}}

\renewcommand{\eqref}[1]{Eq.~(\ref{#1})}

\renewcommand{\tilde}{\widetilde}   % tilde over characters
\renewcommand{\vec}[1]{\mathbf{#1}} % vectors are boldface
\newcommand{\abs}[1]{\ensuremath{\left|#1\right|}}

\renewcommand{\GeV}{\ensuremath{\text{GeV}}}

\renewcommand{\eV}{\ensuremath{\text{eV}}}

\newcommand{\tA}{\ensuremath{\tilde{A}}}

\newcommand{\m}{\ensuremath{\mu}}

\begin{document}

\begin{center}
    {\Large \bf    Time-domain properties of electromagnetic signals in a dynamical axion background }
    
    \vskip 0.5cm
    
%    {\large \it     Casual exponential growth in a causal theory}
       %{\Large  \textsc{for SUSY gauge theories}}
%
%    \vskip 1cm

   \textbf{Peter Adshead}, \textbf{Patrick Draper}, and \textbf{Benjamin Lillard}
 
%    \vspace{-.2cm}
    
    \footnotesize
    \email{adshead@illinois.edu}, \email{pdraper@illinois.edu}, \email{blillard@illinois.edu}

%    \vspace{-.2cm}

%    \begin{institutions}[3.5cm]
%    \footnotesize
%    \textit{University of Illinois at Urbana-Champaign, Department of Physics, Urbana, IL 61801}   
%    \end{institutions}

    \begin{institutions}[3.5cm]
    \footnotesize
    \textit{Illinois Center for Advanced Studies of the Universe \& Department of Physics, 
University of Illinois at Urbana-Champaign, Urbana, IL 61801, USA. }   
    \end{institutions}
\end{center}
%\vspace*{0.5cm}

%\title{Applications of Axion Electrodynamics (Good Paper Title)}
%
%\author{Peter Adshead}
%\email{adshead@illinois.edu} %%ORCID: 0000-0002-5217-7022
%\author{Peter Adshead}
%\email{pdraper@illinois.edu} %%ORCID: 0000-0001-7240-3966
%\author{Yonatan Kahn}
%\email{yfkahn@illinois.edu} %%ORCID: 0000-0002-9379-1838
%\author{Benjamin Lillard}
%\email{blillard@illinois.edu} %%ORCID: 0000-0001-8496-4808
%    \\ \vspace{-.2em}
%    { %\tt
%    \footnotesize
%       \email{adshead@illinois.edu},~
%       \email{pdraper@illinois.edu},~
%       \email{yfkahn@illinois.edu},~
%       \email{blillard@uci.edu}~
%    }

%\affiliation{University of Illinois at Urbana-Champaign, Department of Physics, Urbana, IL 61801}

\date{\today}

\begin{abstract}
Electromagnetic waves in a dynamical axion background exhibit superluminal group velocities at high frequencies and instabilities at low frequencies, altering how photons propagate through space. 
Local disturbances propagate causally, but unlike in ordinary Maxwell theory, propagation occurs inside as well as on the lightcone.  For the unstable  modes, the energy density in the electromagnetic field grows exponentially along timelike displacements. % (inside the lightcone).
In this paper we derive  retarded Green functions in axion electrodynamics in various limits 
and study the time-domain properties of propagating signals. 
%{\bf\color{blue} PD: Double-check whether the following statements are still accurate. Could also consider dropping these sentences; I don't feel strongly.}
%For cosmologically relevant axions, the characteristic timescale for the exponential growth is usually very long (spanning decades or centuries), but the effect might be detected in a sufficiently sensitive terrestrial experiment. 
%In the fuzzy dark matter limit, where the axion mass is so small that its Compton wavelength reaches astrophysical sizes, the relevant timescales can be large enough that the exponential growth of the electric and magnetic fields becomes a dominant effect.  
\end{abstract}
%\maketitle

\setcounter{tocdepth}{2}
\tableofcontents

\section{Introduction}

Axion models provide some of the most well-motivated extensions to the Standard Model, providing a mechanism to resolve the strong $CP$ problem and a class of dark matter candidates.
Through its coupling to gluons, 
%Through its coupling to the %parity ($P$) and charge-parity ($C P$) odd
% $G_{\mu \nu} \tilde{G}^{\mu \nu}$ combination of gluon field strength tensors,
 the vacuum expectation value of the axion field cancels the $\bar\theta$ parameter of quantum chromodynamics (QCD) and restores $CP$ symmetry~\cite{Peccei:1977hh,Peccei:1977ur}, explaining the surprisingly small experimentally measured value, $\abs{\bar\theta} < 6\times 10^{-11}$~\cite{Baker:2006ts,Afach:2015sja}. %, consistent with the preservation of $CP$ symmetry by the strong interaction.
Through the misalignment mechanism~\cite{Abbott:1982af,Preskill:1982cy,Dine:1982ah}, axions can also be produced in the early universe in sufficient abundance to comprise most or all of the dark matter.
%Axions generically have feeble couplings to the Standard Model particles, making them viable dark matter models for a wide range of parameters.
%For small enough masses

Many of the most stringent constraints on  axion models utilize a  coupling between the axion and the Standard Model electric and magnetic fields,
\begin{align}
\mathcal L \supset - \frac{1}{8} g_{a \gamma\gamma} a \epsilon^{\mu\nu\rho\sigma} F_{\mu \nu} F_{\rho\sigma} = g_{a \gamma\gamma} a \vec{E} \cdot \vec{B}.
\label{eq:EMcoupling}
\end{align}
This coupling enables axion production through the Primakoff process~\cite{Primakoff:1951pj,Sikivie:1983ip}; axion decay to photons; and axion--photon interconversion in the presence of electromagnetic fields. As such it can be used either to detect or to produce axions in the laboratory.
%Detecting the existence of axions directly through their coupling to QCD is challenging,
%although proposals that make use of nuclear magnetic resonance~\cite{Budker:2013hfa, JacksonKimball:2017elr,Geraci:2017bmq} may produce results in the near future.
%Axions can be created from photons in the presence of an external electromagnetic field
%or in stellar environments through the Primakoff process~\cite{Primakoff:1951pj}.
%This coupling 
%Stellar cooling via the Primakoff process is efficient 
%\cite{Dicus:1979ch,Raffelt:2006cw}
Stellar observations~\cite{Dicus:1979ch,Raffelt:2006cw, Anastassopoulos:2017ftl} constrain $g_{a \gamma\gamma} < 10^{-10}\, \GeV^{-1}$ for a wide range of axion masses,
and additional constraints set by the power spectra of bright X-ray point sources provide a more stringent limit of $g_{a\gamma\gamma} \lesssim 10^{-12}\, \GeV^{-1}$ for light axions of mass $m_a \lesssim 10^{-12}\, \eV$~\cite{Brockway:1996yr,Grifols:1996id,Conlon:2017qcw}.
%For sufficiently light axions, when the axion de~Broglie wavelength is macroscopically large,
At large occupation numbers and de Broglie wavelengths, axion dark matter behaves as a classical, oscillating background field
%with a coupling $\mathcal L \sim a \vec{E} \cdot \vec{B}$ to electric and magnetic fields.
that induces small time-dependent perturbations to  electrodynamics,
which can be probed with a variety of different sensitive experimental techniques~\cite{Sikivie:1983ip,Du:2018uak,Ouellet:2018beu,Zhong:2018rsr,Armengaud:2019uso,Berlin:2019ahk,Lasenby:2019prg} .

The coupling in \eqref{eq:EMcoupling} also affects the propagation of classical electromagnetic radiation~\cite{Wilczek:1987mv,Carroll:1989vb,Coriano:1992bh}. 
Electromagnetic plane waves traveling through an axion background acquire modified phase velocities for left- and right-polarizations, an effect which may be observable in interferometers~\cite{DeRocco:2018jwe,Obata:2018vvr}, atomic clocks~\cite{Krauss:2019lqo} or astrophysical sources~\cite{Chigusa:2019rra}, for some ranges of axion masses and couplings.  Furthermore, low frequency modes exhibit tachyonic instabilities, while at high frequencies, group velocities for both polarizations are superluminal.
%Causality is preserved in the theory 

Despite the presence of plane wave solutions with superluminal group velocities, axion electrodynamics is a causal theory: local disturbances do not propagate outside the lightcone. This was first shown long ago in the case of tachyonic scalar field theory by Aharonov, Komar, and Susskind~\cite{Aharonov:1969vu}. Here we show that the same mechanism is at work in axion electrodynamics, %with interesting consequences for the time-domain evolution of local disturbances.
with consequences that include the exponential growth of local disturbances.

%\bl{Consider citing~\cite{Andrianov:1998wj,Kostelecky:2000mm,Adam:2001ma,Frank:2006ww,Kharlanov:2009pv, Alfaro:2009mr,Blas:2019qqp,  Sikivie:2020zpn}}

\medskip

In this paper we calculate the classical electromagnetic retarded Green function %to arbitrary order in $g_{a \gamma \gamma}$ 
in a coherent, dynamical axion background in several disparate regimes of axion parameter space.
Our results are organized based on the hierarchical ordering of three different scales:
\begin{itemize}
\item $m_a$, the axion mass;
\item $\mu_0 \equiv \frac{1}{2} g_{a\gamma\gamma} \sqrt{\rho_a}$, a mass scale that determines the rate of exponential growth of the electromagnetic fields, based on the axion density $\rho_a$; and
\item $T^{-1}$ and $L^{-1}$, the inverses of the characteristic  propagation time and distance $T\sim L$ of a signal.
\end{itemize}
Rather than focusing only on those axion models that provide a natural solution to the strong $CP$ problem, we consider the broader realm of axion-like particles (ALPs), where the axion mass $m_a$ and decay constant $f_a$ are not required to satisfy $m_a f_a \sim m_\pi f_\pi$, and the value of $g_{a\gamma\gamma}$ is not determined by $m_a$. These axions can still provide a wide range of dark matter candidates (see, e.g.,~\cite{Arias:2012az, Blinov:2019rhb, DiLuzio:2020wdo}). For the lowest-mass ``fuzzy dark matter'' candidates, generic constraints on ultra-light scalars from Lyman-$\alpha$ data~\cite{Armengaud:2017nkf,Irsic:2017yje,Kobayashi:2017jcf,Nori:2018pka} impose a lower bound on the axion mass of $m_a \gtrsim 2 \times 10^{-21}\, \eV$, though in the ``large misalignment'' regime of Ref.~\cite{Arvanitaki:2019rax} the Lyman-$\alpha$ bound is altered by the effect of ALP self-interactions.
Our analysis in this paper encompasses the nearly twenty decades of ALP parameter space above this bound, where the axion is still light enough that it can be treated as a coherently oscillating background field. 

%{\bf\color{blue} PD: comment here explicitly that our analysis is only in the $\nabla a=0$ limit?}
%
In terms of the axion virial velocity $v$, the mass sets an upper bound on the characteristic $L$ and $T$,  $L \lesssim (m_a v)^{-1}$ and $T \lesssim (m_a v^2)^{-1}$, after which any analysis must incorporate the effects of decoherence.
In the example of fuzzy dark matter with $m_a \approx 2\times 10^{-21}\, \eV$, and taking $v \sim 10^{-3}$ as the ALP virial velocity, the coherence length and time are respectively $L_c \lesssim 10^{ 17}\, \text{m} \approx 3\,\text{pc}$
and $T_c \lesssim 3\times 10^{11}\, \text{s} \approx 10^4\,\text{yr}$.
On the other extreme, for $m_a$ much larger than $10^{-4}\,\eV$, even a tabletop experiment will encounter significant decoherence. 
Our analysis is focused on the nonrelativistic limit, neglecting these decoherence effects and ignoring spatial gradients in the axion field.
%Our treatment of the axion as a coherent field is valid only for $L$ and $T$ smaller than the coherence length and time, respectively. 

\medskip

%{\bf\color{blue} PD: make positive comments below about the interesting time-domain properties / propagation inside the lightcone, echoing the abstract. }

We begin in Section~\ref{sec:static} with the simplest analysis, the $m_a \ll \mu_0, 1/T$ limit. In this case the value of the axion field changes at an approximately constant rate, $\partial_t a(x, t) \approx \textit{const}$, and we find an analytic solution to the Green function valid for all values of $\mu_0 T$.
This Green function exhibits exponential growth inside the lightcone of the disturbance when $\mu_0 T \gtrsim 1$. 
Despite the potentially catastrophic consequences of this unbounded growth, the dilute density of dark matter and experimental constraints on the axion--photon coupling ensure that the timescales for the genuinely exponential phase of the growth are outside the reach of all but the lightest ALP candidates, unless the local ALP density $\rho_a$ is enhanced by several orders of magnitude above $0.4\, \GeV/\text{cm}^3$.
In Section~\ref{sec:constnumerics} we highlight some of the curious and potentially detectable perturbations to classical electrodynamics induced by the axion background.

For almost all allowed values of $g_{a\gamma\gamma}$ and $\rho_a$, the hierarchy $\mu_0 \ll m_a$ is more realistic, and we explore this limit in Section~\ref{sec:osc}. 
In the case of the oscillating background axion field it is no longer possible to derive an exact analytic expression for the Green function using the methods of Section~\ref{sec:static}. 
Instead, we construct perturbative expansions for the  $\mu_0 T \ll 1$ and  $\mu_0 T \gg 1$ limits by expressing the Green function as a continued fraction.
When the frequency support of the radiation includes $\omega \approx \frac{1}{2} m_a$, a narrow resonance induces exponential growth for large $T$.
In Section~\ref{sec:resonant} we calculate the dominant part of the Green function  in the $\mu_0 T \gg 1$ limit. In this late-time limit the resonant enhancement dwarfs the contribution from frequencies $\omega \neq \frac{1}{2} m_a$. 
This resonant emission has been previously studied in~\cite{Arza:2018dcy,Hertzberg:2018zte}, although the resonant band is so narrow that dispersion effects and gravitational redshifting may completely prevent the exponential growth~\cite{Arza:2020eik}.
For the non-resonant limit $\mu_0 T \ll 1$, and for electromagnetic signals which do not include support near the resonant frequency $\omega \approx \frac{1}{2} m_a$, Section~\ref{sec:notresonant} provides a continued fraction expression for the Green function that is valid to arbitrary order in $g_{a\gamma\gamma}$.
In Sections~\ref{sec:constnumerics}~and~\ref{sec:oscnumeric}, we provide numeric examples to illustrate the behavior of signals in various corners of ALP parameter space, and to verify our analytic expressions.

The primary results of this paper are collected in Eqs.~(\ref{eq:greensoln2d}),~(\ref{eq:corrected}) and~(\ref{eq:continued2}) in Section~\ref{sec:conclusion}.
Despite the significant differences between the two limits, the Green functions of Sections~\ref{sec:static} and~\ref{sec:osc} both exhibit the novel inside-the-lightcone propagation and exponential growth in certain modes.

%{\bf\color{blue} PD: maybe say something like ``the primary results of this paper are collected in Eq ... in the conclusions"}

%In this work we use the mostly-minus metric convention, $\eta^{\mu \nu} = \text{Diag}(+1, -1, -1, -1)$, and natural units $\hbar  = c = 1$.

\subsection{Axion Electrodynamics}

In terms of $\theta(x^\mu)$, the local value of the effective CP violation induced by the axion background, the Lagrangian for electrodynamics includes the interactions
\begin{align}
\mathcal L = - \frac{1}{4} F_{\mu \nu} F^{\mu \nu} - A_\mu J^\mu +  \frac{ \theta }{8}  \epsilon^{\mu\nu\rho\sigma} F_{\mu \nu} F_{\rho \sigma}
\end{align}
where $F_{\mu \nu}$ is the electromagnetic field strength tensor, $A_\mu$ and $J^\mu$ are the vector potential and 4-current, %and $\theta(x^\mu) \equiv g_{a \gamma \gamma} a(x^\mu)$ is the local value of the effective CP violation induced by the axion background.
and  $\theta(x^\mu)$ is related to the value of the axion field via
\begin{equation}
\theta(x^\mu) \equiv g_{a \gamma \gamma} a(x^\mu).
\end{equation}
In Lorenz $\partial_\alpha A^{\alpha} = 0$ gauge, the equations of motion for $A_\mu$ reduce to
\begin{align}
\partial^2 A^\mu  -  \epsilon^{\mu \nu \rho \sigma} (\partial_\nu \theta) (\partial_\rho A_{\sigma} ) = J^\mu ,
\end{align}
which depends explicitly on the derivatives of $\theta(x^{\mu})$ rather than $\theta$ itself. 
Taking the external source to be neutral and transverse, $J^0  = 0$ and $\nabla \cdot J = 0$,
and neglecting any spatial gradients in the background axion field, $\abs{\nabla a} \ll \abs{\dot a}$,
 the equations of motion for the scalar and vector potentials decouple,\footnote{The equations of motion with $\nabla a \neq 0$ are discussed in e.g.~\cite{Carroll:1989vb,Sikivie:2020zpn}.}
\begin{align}
\partial^2 \Phi = 0,
&&
\partial^2 \vec{A} + \dot\theta \nabla\times \vec{A} = \vec{J}.
\label{eq:AEDeom}
\end{align}
%For concreteness, let us consider plane waves propagating in the $\hat{z}$ direction, with the 
For a typical model of ALP dark matter, $\dot\theta (t)$ is given by
\begin{align}
\theta(t) \approx \theta_0 \cos(m_a t),
&&
\theta_0 = g_{a \gamma \gamma} \frac{\sqrt{\rho_a} }{m_a},
&& \dot \theta_0 = m_a \theta_0 %= g_{a\gamma\gamma} \sqrt{\rho_\chi}
\label{eq:theta0gagg}
\end{align}
where the value of $\theta_0$ is set by the local axion dark matter density, $\rho_a \sim (0.042\, \eV)^4$, and where
%Given the constraints on $g_{a \gamma \gamma}$, the mass scale $\dot \theta$ is usually quite small.
\begin{equation}
\dot\theta_0 \simeq 1.75\cdot 10^{-23}\, \eV \times \left( \frac{g_{a\gamma\gamma} }{10^{-11}\, \GeV^{-1}} \right) \sqrt{ \frac{\rho_a}{0.4 \, \GeV/\text{cm}^3} }.
\label{eq:dottheta}
\end{equation}
For future reference, we note that
$
1.75\cdot 10^{-23}\, \eV
\simeq 2.66\cdot 10^{-8} \, \text{Hz} 
\simeq 0.84\, \text{yr}^{-1}
$.

\section{Green Functions for the Steady-State Background}
\label{sec:static}
%
% In the simplest canonical constructions~\cite{Peccei:1977ur,Kim:1979if,Shifman:1979if,Zhitnitsky:1980tq,Dine:1981rt}, the axion is a pseudo-Nambu--Goldstone boson associated with the spontaneous breaking of an approximately conserved $U(1)_\text{PQ}$ global symmetry at a scale $f_a$.
%This \upq\ is also explicitly broken by nonperturbative effects at a characteristic scale $\Lambda$, 
%generating a mass for the axion of order $m_a \sim \Lambda^2/f_a$.
%In the case of the QCD axion it is the chiral anomaly with the strong interaction that breaks \upq, so that $\Lambda$ is set by $\Lambda_\text{QCD} = \sqrt{m_\pi f_\pi} \sim 100~\MeV$, but in more generic ALP models both $\Lambda$ and $f_a$ are free parameters.
%
%While the precise value of the coefficient $g_{a \gamma\gamma}$ is a model dependent quantity, it is generically non-zero:
%\begin{equation}
%g_{a\gamma\gamma} = \frac{\alpha}{2\pi f_a} \left( \frac{E}{N} - \frac{2}{3}\, \frac{4+z}{1+z} \right),
%\end{equation} 
%where $f_a$ is the axion decay constant, $E$ and $N$ represent the electromagnetic and color anomaly coefficients of the Peccei--Quinn current, and $z = m_u/m_d \approx 0.5$~\cite{Tanabashi:2018oca}. 
%For the standard KSVZ~\cite{Kim:1979if,Shifman:1979if} and DFSZ~\cite{Zhitnitsky:1980tq,Dine:1981rt} axion models, this ratio is $E/N = 0$ and $E/N = 8/3$, respectively, although in principle $g_{a \gamma \gamma}$ may be enhanced or suppressed by changing the charged fermion content of the high-energy theory.

For timescales that are short compared to the period of the axion oscillation, $m_a t \ll 1$ and $m_a \lesssim \dot\theta$, it is appropriate and instructive to consider the approximation $\partial_t^2 \theta \approx 0$, where $\dot\theta(t)$ assumes a nearly constant value $-\dot\theta_0 \leq \dot\theta(t) \leq \dot\theta_0$.
In this case the differential equation for $\vec{A}$ can be solved using a Fourier transform,
\begin{align}
\vec{A}(t, \vec{x}) &= \left( \frac{1}{2\pi} \right)^4 \int d\omega d^3 \vec{k}\; \tilde{ \vec{A}}(\omega, \vec{k} ) e^{i (\omega t - \vec{k} \cdot \vec{x} ) } , 
%\\
%\partial_i A_j &= \left( \frac{1}{2\pi} \right)^4 \int d\omega d^3 \vec{k}\; \tilde A_j(\omega, \vec{k} ) ( - i k_i) e^{i (\omega t - \vec{k} \cdot \vec{x} ) },
\end{align}
so that the differential equation for $\vec{A}$ becomes a set of algebraic equations for $\tilde A_i$.
For planar waves propagating in the $\hat{z}$ direction, $\vec{k} = k \hat{z}$, the polarization basis $\vec{A} = A_+ \hat{\epsilon}_+ + A_- \hat{\epsilon}_- + A_z \hat{z}$
with $\hat\epsilon_\pm = \frac{ 1}{\sqrt{2}}( \hat{x} \pm i \hat{y})$
diagonalizes the equations of motion,
%In the coordinate basis where $\vec{k} = k \hat{z}$ is in the $\hat{z}$ direction, the equations of motion are diagonalized  for circularly polarized radiation, $\vec{A} = A_+ \hat{\epsilon}_+ + A_- \hat{\epsilon}_- + A_z \hat{z}$,
%where
%\begin{equation}
%%A_\pm = \frac{ A_x \pm i A_y }{\sqrt{2}},
%\epsilon_\pm = \frac{ \hat{x} \pm i \hat{y} }{\sqrt{2}},
%\end{equation}
with the result
\begin{align}
\tA_\pm \left[ k^2 - \omega^2 \pm  \dot\theta k \right] =&\; \tilde{j}_\pm(\vec{k}), &
\tA_z \left[ k^2 - \omega^2  \right] =&\; 0 , %&
%A_\pm = \frac{ A_x \pm i A_y }{\sqrt{2}},
\label{eq:eom}
\end{align}
%where
%\begin{equation}
%A_\pm = \frac{ A_x \pm i A_y }{\sqrt{2}},
%%\epsilon_\pm = \frac{ \hat{x} \pm i \hat{y} }{\sqrt{2}},
%\end{equation}
where $\tilde{j}_\pm(k)$ is the Fourier transform of the transverse source $(\hat{\epsilon}_{\pm} \cdot \vec{J})$ in the polarization basis.
Circularly polarized plane waves propagate with the dispersion relations
\begin{equation}
\omega_\pm^2 = k^2 \pm k \dot\theta, %\equiv k^2 \pm k \Theta,
\label{eq:dispersion}
\end{equation}
%where we have defined the constant $\Theta \equiv \alpha \dot \theta/\pi$ for convenience.
producing subluminal or superluminal phase velocities depending on the sign of $\dot\theta$ and the polarization of the radiation. %\bl{Define $\frac{\alpha \dot\theta}{\pi} = \Theta$?}

The group velocities for both modes are superluminal, for both positive and negative $\dot\theta$,
\begin{equation}
\frac{d \omega_{\pm} }{d k} %= \left( k^2 \pm k \Theta \right)^{-\frac{1}{2} } \left( k \pm \frac{\Theta}{2}\right)
= \frac{k \pm \frac{1}{2} \dot\theta }{\sqrt{ k^2 \pm k \dot\theta} } \geq 1,
\end{equation}
with $d\omega_\pm / dk = 1$ only for $\dot\theta = 0$.
The effect appears at quadratic order in $ \dot\theta/k$, % $\alpha \dot\theta/k \ll 1$ shows 
\begin{align}
\frac{d \omega_{\pm} }{d k} &= 1 + \frac{1}{8} \frac{ \dot\theta^2}{ k^2} \mp \frac{1}{8} \frac{\dot\theta^3}{ k^3}  + \ldots %\mathcal O\left(\left( \tfrac{\alpha \dot\theta}{\pi k}\right)^4\right).
\end{align}
so that for $\dot\theta \ll k$ the modification of the group velocity is subdominant to the $\mathcal O( \dot\theta/k)$ change in the phase velocity,
\begin{align}
\frac{\omega_\pm}{k} = \sqrt{1 \pm \frac{\dot\theta}{k}} = 1 \pm \frac{1}{2} \frac{\dot\theta}{k} -\frac{1}{8} \frac{\dot\theta^2}{k^2} \pm \frac{1}{16} \frac{\dot\theta^3}{k^3} + \ldots .
\label{eq:phasev}
\end{align}

We demonstrate below that the retarded Green function vanishes outside the light cone, preserving causality despite the presence of superluminal group velocities. As a necessary consequence, disturbances in the field induced by local sources grow exponentially in timelike directions. 
%In Section~\ref{sec:green2d} we show this explicitly by calculating the Green function.
To leading order in $\dot\theta$, the phase velocities alternate about a central value $\omega_{\pm} / k = 1$ based on the polarization of the light and the sign of $\dot\theta$. After multiple periods of the axion oscillation, the perturbations to the phase velocity tend to cancel each other.
On the other hand, the group velocity is superluminal for both positive and negative $\dot\theta$, so the exponential growth is not ameliorated by any periods of exponential decay when the sign of $\dot\theta$ changes. The effects from the modified group velocities should grow over time.

With the approximation that $\dot\theta(t)$ is nearly constant, the two-dimensional Green function can be obtained analytically  to all orders in $\dot\theta$. 
Experimental constraints on $g_{a\gamma\gamma} < 10^{-10}\, \GeV^{-1}$ and $m_a$ exclude the $m_a \ll \dot\theta$ possibility unless the local axion density is significantly enhanced, $\rho_a \gg \mathcal O( \GeV/\text{cm}^3)$, so the results in this section are directly applicable  primarily to situations involving to dense clumps of ultralight axions. 
When we calculate the Green function for the more broadly relevant $m_a \gg \dot\theta$ hierarchy of scales in Section~\ref{sec:osc}, the steady-state case with constant $\dot\theta$ also provides a helpful consistency check in the limit where the exponential growth becomes important.

\subsection{Green Function Solution in Two Dimensions} \label{sec:green2d}

The Green function can be calculated analytically for the simplified case of plane waves $\vec{k} = k \hat{z}$ with a spatially homogenous ($\nabla \theta = 0$), steady-state ($\partial^2_t \theta = 0$) axion background. Imposing translational symmetry in $x$ and $y$ effectively reduces the system from $(3+1)$ dimensions to $(1+1)$.
%For $\dot \theta (t) \sim \dot\theta_0 \sin(m_a t)$, the resulting equations of motion invoke the Mathieu function.
%Before we consider this more generic behavior, it is instructive to begin with the further approximation that $\partial_t^2 a = 0$, so that $\dot\theta$ is treated as a constant. 
%
%% and $\partial_t^2 \theta =0$.
%%Sentence about it
%For a wave propagating in the $\vec{k} = k \hat{\vec{z}}$ direction,
%the polarization basis
%\begin{equation}
%A_\pm = \frac{ A_x \pm i A_y }{\sqrt{2}}
%\end{equation}
%diagonalizes the equations of motion, with dispersion relations for $A_\pm$ given by
%\begin{align}
%\omega_\pm^2 = k^2 \pm k  \dot\theta ,
%\end{align}
%where $\omega_\pm$ is the frequency of the right- or left-polarized light, and $k$ is its momentum in the $\hat{z}$ direction.
The $\tA_\pm$ equation of motion in \eqref{eq:eom} admits a Green function solution $g_{\pm}(z, t)$ of the form
\begin{align}
%J_\pm^\mu &= \int\! d\tau\, d\lambda\, s_\pm(\tau, \lambda) \delta(z - \lambda) \delta(t- \tau) \\
A_\pm &= \int\! dt_0\, dz_0\, j_\pm(t_0, z_0) g_\pm(t - t_0, z - z_0),
\label{eq:greendef}
\end{align}
where
\begin{equation}
(\partial_t^2 - \partial_z^2 \pm i \dot\theta \partial_z)  g_\pm(t - t_0, z - z_0) = \delta(z - z_0) \delta(t - t_0).
\label{eq:greenDE}
\end{equation}
In this section it is convenient to fold a factor of $1/2$ into the definition of $\dot\theta$,
\begin{equation}
\m(t) = \frac{ \dot\theta(t) }{2 },
\label{eq:em}
\end{equation}
where $\m$ determines the rate of exponential growth, as we show below.

Defining a related Green function $G_0$,
\begin{equation}
g_{\pm}(t|t_0, z|z_0) = e^{\pm i \m z} G_0(t|t_0, z|z_0),
\end{equation}
%and folding a factor of $\alpha/2\pi$ into the definition of $\dot\theta$,
%\begin{equation}
%\m(t) \equiv \frac{\alpha \do\theta(t) }{2 \pi},
%\end{equation}
\eqref{eq:greenDE} can be simplified to
\begin{equation}
 \left( \partial_t^2 - \partial_z^2 - \m^2 \right) G_0(t, z) = e^{\mp i \m z} \delta(z) \delta(t) ,
\label{eq:G0eom}
\end{equation}
so that $\m^2$ acts as an effective tachyonic mass for the scalar-like Green function $G_0$.

Applying the Fourier transform and integrating both sides of \eqref{eq:G0eom} produces the integral form of the Green function,
\begin{align}
G_\epsilon = \int \frac{d\omega dk}{(2\pi)^2} \frac{e^{-i(\omega t - k z) } }{k^2 - (\omega+ i \varepsilon)^2 - \m^2},
\label{eq:G0int}
\end{align}
where $\varepsilon >0$ indicates that the contour in the complex $\omega$ plane should correspond to the retarded Green function, which vanishes for $t < 0$.
For $k^2 > \m^2$ the $\varepsilon \rightarrow {0^+}$ limit can be recovered easily. However, for $k^2 < \m^2$ one of the poles in $\omega$ is located above the real axis, at
\begin{equation}
\omega = -i \varepsilon \pm i \sqrt{ \m^2 - k^2} .
\end{equation}
To recover the retarded Green function, the contour in $\omega$ should pass above both poles, with $\varepsilon \rightarrow {\mu^+}$ on the imaginary axis.

\begin{figure}[t]
\centering
\includegraphics[width=0.7\textwidth]{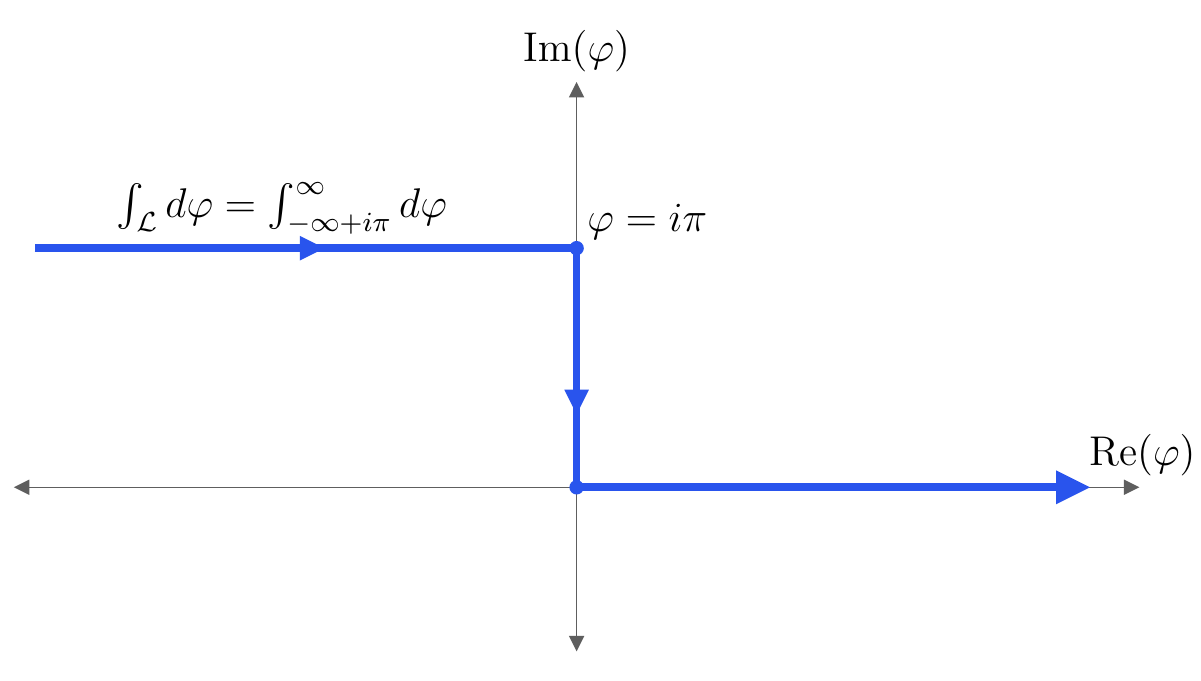}
%\begin{tikzpicture}
%	\node[anchor=south west,inner sep=0] (plot) at (0,0) {\includegraphics[width=0.68\textwidth]{plots/contour.pdf}};
%	\begin{scope}[x={(plot.south east)},y={(plot.north west)}]
%		\node at (0.25, 0.75) {$\int_\mathcal L d\varphi = \int_{-\infty + i\pi}^\infty d\varphi$};
%		\node (ylabel) at (0.5, 1.0) {$\text{Im}(\varphi)$};
%		\node (xlabel) at (0.98, 0.36) {$\text{Re}(\varphi)$};
%		\node (pi) at (0.56, 0.73) {$\varphi = i \pi$};
%	\end{scope}
%\end{tikzpicture}	
\caption{The contour $\mathcal L$ that allows \eqref{eq:GepsilonL} to be written in terms of Hankel functions. The shorthand notation for this contour is $\int_{-\infty + i\pi}^\infty d\varphi$, indicating that the contour approaches the imaginary axis from negative real $\infty$ with a constant imaginary component of $+i \pi$.
}
\label{fig:contour}
\end{figure}

%With some coordinate substitutions and integration along a particular contour, it is possible to find an analytic solution for the integral in \eqref{eq:G0int} in terms of Bessel functions.
The $\int d\omega$ integral can be completed using the residue theorem,
\begin{equation}
G_\epsilon = \Theta(t) \frac{-2\pi i e^{-\varepsilon} }{(2\pi)^2 } \int \frac{dk\, e^{i k z} }{2 \omega_0} \left( e^{i \omega_0 t } - e^{-i \omega_0 t} \right),
\end{equation}
where $\omega_0 = \sqrt{k^2 - \m^2 } = i \sqrt{\m^2 - k^2}$ and $\Theta(t)$ is the  step function.
The integral is simplified by a coordinate substitution $k \rightarrow \varphi$,
\begin{align}
k = \m \cosh \varphi , &&
\omega_0 = \m \sinh \varphi ,
\label{eq:magic}
\end{align}
where the contour $\mathcal L$ in the complex $\varphi$ plane
is shown in Figure~\ref{fig:contour}, and allows $\cosh \varphi$ to vary smoothly from $-\infty$ to $\infty$ with $\text{Im}(\cosh \varphi ) = 0$.
% must be such that $-\infty < \cosh \varphi < \infty$. This is achieved by
%$\varphi = x + i \pi$ for real $x < 0$; then, with $\varphi = i (\pi - y)$ for $0 \leq y < \pi$; then, $\varphi = x$ for $x \geq 0$.
%Along this contour, $\cosh \varphi$ varies smoothly from $-\infty$ to $\infty$ with $\text{Im}(\cosh \varphi ) = 0$, and
%$\sinh \varphi$ obeys the prescription we have specified for $\omega_0$.
% and $\Theta(t)$ is the Heaviside step function.
%By defining a coordinate substitution $k \rightarrow \varphi$
%\begin{align}
%k = \m \cosh \varphi &&
%\omega_0 = \m \sinh \varphi
%\end{align}
%and 
By mapping the coordinates $(t, z)$ to  $(\eta, \lambda)$ via
\begin{align}
\m t =  \sqrt{\lambda} \cosh \eta , &&
\m z = \sqrt{\lambda} \sinh \eta , &&
t^2 - z^2 =  \lambda/\mu^2,
\label{eq:lambda}
\end{align} 
for $\lambda \geq 0$,
the integral can be written in terms of just $\lambda$ and $\varphi\pm \eta$:
\begin{align}
G_0(\lambda \geq 0) &= \frac{ \Theta(t)}{4\pi i} \int_{\mathcal L} d\varphi \left( e^{i \sqrt{\lambda} \sinh (\varphi + \eta) } - e^{-i \sqrt{\lambda} \sinh(\varphi - \eta) } \right)
\nonumber\\
& = \frac{ \Theta(t)  }{4\pi i} \int_{-\infty + i\pi}^\infty d\varphi'  \left( e^{i \sqrt{\lambda} \sinh \varphi' } - e^{-i \sqrt{\lambda} \sinh \varphi' }  \right).
\label{eq:GepsilonL}
\end{align}
%where the further coordinate substitution $\varphi_\pm \equiv \eta \pm \varphi$ shifts the $\eta$ dependence from the integrand into the contour, $\mathcal L \rightarrow \mathcal L_\pm$.
%Noting that the contours $\mathcal L_+$ and $\mathcal L_-$ meet only at the point $\varphi = \eta$, and that the two integrands are identical, the integral $G_0$ can be rewritten in terms of a $\mathcal L_<$ and a $\mathcal L_>$, so that $\mathcal L_{<, >}$ include $\text{Re}(\varphi) \leq \eta$ and $\text{Re}(\varphi) \geq \eta$, respectively:
%\begin{align}
%G_0 = \frac{i \Theta(t)  }{4\pi} \left( \int_{\mathcal L_<} d\varphi e^{i \sqrt{\lambda} \sinh \varphi } +  \int_{\mathcal L_>} d\varphi e^{i \sqrt{\lambda} \sinh \varphi }  \right).
%%\label{eq:GepsilonL}
%\end{align}
%In terms of $\varphi = x + iy$ for real $x$ and $y$, the contour $\mathcal L_<$ follows $\varphi = x + i\pi$ for $-\infty < x < \eta$; then $\varphi = \eta + i y$ from $y=\pi$ to $y = 0$; then returns to $-\infty$ along the real axis, with $\varphi = x$ for $\eta \geq x$.
%The contour $\mathcal L_>$ can be found by rotating $\mathcal L_<$ about the point $\varphi = \eta$ by $180^\circ$.
Here the notation $\int_{-\infty + i \pi}^\infty d\varphi$ indicates
the imaginary offset for $\text{Re}(\varphi) < 0$ shown in Figure~\ref{fig:contour}.
% that the contour runs parallel to the real axis with $\varphi = x + i \pi$ for $-\infty < x \leq 0$; follows the imaginary axis as $\varphi = i y$ for $\pi \geq y \geq 0$; then follows the positive real axis, $\varphi = x$ with $0 \leq x < +\infty$, as shown in Figure~\ref{fig:contour}. 
%This $\mathcal L_+$ runs parallel to the real axis with $\text{Im}(\varphi) = + i \pi$ from $-\infty < \text{Re}(\varphi) \leq \eta$; follows $\varphi = \eta + i y$ for $\pi \geq y \geq 0$; then runs along the real axis as $\varphi = x$ for $\eta \leq x < + \infty$. As there are 
In order to make this last simplification, removing the $\eta$ dependence completely, note that the integrand has no poles for finite $\varphi'$, so that the contours $\mathcal L_\pm$ from the coordinate substitutions $\varphi_\pm = \eta \pm \varphi$ can be shifted horizontally to compensate for $\eta$.

Outside the lightcone, for spacelike displacements $z^2 > t^2$, the coordinate transformation \eqref{eq:lambda} is replaced by the alternative
\begin{align}
\m z = \bar z \cosh \eta , &&
\m t =  \bar z \sinh \eta , &&
z^2 - t^2 = \bar z^2 = - \lambda/\mu^2.
\label{eq:zbar}
\end{align} 
In this case, with $\lambda <0$, the two contributions to the integral cancel each other, 
\begin{align}
G_0(\lambda < 0) &= \frac{ \Theta(t)}{4\pi i} \int_{\mathcal L} d\varphi \left( e^{i \bar z \cosh (\varphi + \eta) } - e^{i \bar z \cosh(\varphi - \eta) } \right) = 0,
\end{align}
and so the retarded Green function vanishes outside the lightcone.
%\\
%G_0(\lambda) &= \frac{ \Theta(t) \Theta(\lambda) }{4\pi i} \int_{\mathcal L} d\varphi \left( e^{i \sqrt{\lambda} \sinh (\varphi + \eta) } - e^{-i \sqrt{\lambda} \sinh(\varphi - \eta) } \right).
%\end{align}

Using this contour notation, and the integral definition of the Hankel functions % $H_\nu^{(1,2)}$,
\begin{equation}
H^{(1,2)}_\nu (z) = \pm \frac{1}{\pi i} \int_{-\infty}^{+\infty \pm i \pi}\!du\, e^{z \sinh u - \nu u},
\label{eq:hankel}
\end{equation}
 $G_0$ can be written as
\begin{align}
G_0(t,z) = \frac{\Theta(t)\Theta(\lambda) i}{4\pi} \left( \frac{\pi}{i} H^{(2)}_0 (i \sqrt{\lambda}) - \frac{\pi}{i} H^{(2)}_0 (-i \sqrt{\lambda})  \right).
\label{eq:g0penultimate}
\end{align}
\eqref{eq:g0penultimate} simplifies for integer values of $\nu$ to recover an expression in terms of  the %exponentially increasing 
modified Bessel function of the first kind, $I_\nu(z)$:
\begin{align}
G_0(t,z) = \frac{\Theta(t) \Theta(\lambda)}{2} I_0(\sqrt{\lambda}).
\end{align}
Finally, we obtain
\begin{align}
%\label{eq:g0ultimate}
%\end{align}
%so that the retarded Green function of \eqref{eq:greendef} is found to be
%\begin{equation}
g_\pm(t, z) = e^{\pm i \m z} \frac{\Theta(t) \Theta(t^2 - z^2) }{2}  I_0 \left( \sqrt{\m^2 (t^2 -  z^2)} \right).
\label{eq:greensoln2d}
\end{align}
%\paragraph{Asymptotic Exponential Growth:}

\medskip

As anticipated by the discussion of tachyonic scalar theories in Ref.~\cite{Aharonov:1969vu}, 
the Bessel function $I_0(\sqrt{\lambda})$ grows exponentially with timelike $\lambda \propto t^2 - z^2$. In the $\lambda \gg 1 $ limit it approaches
\begin{equation}
\lim_{\lambda \rightarrow \infty} I_0(\sqrt{\lambda} ) = \frac{\lambda^{-1/4} e^{\sqrt{\lambda}} }{\sqrt{2\pi} } \left( 1 + \frac{1}{8\sqrt{\lambda}} + \mathcal O (\lambda^{-1}) \right).
\label{eq:besselasymptotic}
\end{equation}
%Not only does the Green function vanish outside the light cone, $t^2 < z^2$, but it explicitly satisfies the equations of motion,
The retarded Green function vanishes outside the light cone, $t^2 < z^2$, and it explicitly satisfies the equations of motion
\begin{equation}
\int\! dt_0\, dz_0\, j_\pm(t_0, z_0) \left(\partial_t^2 - \partial_z^2 \pm i \dot\theta  \partial_z \right)  g_\pm( t - t_0, z - z_0) = j_\pm(t, z).
\end{equation}
%\newpage
From the  solution for $A_\pm$ one can derive the corresponding Green functions for the $\vec{E}$ and $\vec{B}$ fields,
\begin{align}
E_x = - \partial_t A_x
&&
E_y = \partial_t A_y,
&&
B_x = - \partial_z A_y,
&&
B_y = \partial_z A_x,
%&&
%E_z = B_z = 0,
\end{align}
with $E_z = B_z = 0$.

\medskip

In many of the simplest cases of interest, including idealized axion interferometers and astrophysical sources, the photon source is localized in space and produces a signal $s(\tau, z) = \delta(z) s(\tau)$ or alternatively $s(\tau, z) = s(\tau - z)$ that varies in time and propagates in the forward direction, $z \geq 0$.
With this choice, the $\vec{E}$ and $\vec{B}$ fields can be found from
the $\Theta(z)$ components of the derivatives of $A_\pm$,
and the familiar Standard Model limit $\mu \rightarrow 0$ is recovered by
\begin{align}
\partial_t A^{\mu=0}_\pm(t, z) &= -\Theta(z) \frac{1 }{2} j_\pm(t-z) , \\
\partial_z A^{\mu=0}_\pm(t, z) &= \Theta(z) \frac{1 }{2} j_\pm(t-z) .
\end{align}

%
%%\begin{widetext}
%\begin{align}
%\partial_t A_\pm(t, z) &= \Theta(z) \frac{e^{\pm i \m z } }{2} \left( j_\pm(t-z) + \int_{-\infty}^{t-z} \! dt_0\, j_\pm(t_0) \frac{\m^2(t - t_0) I_1( \sqrt{\lambda'} ) }{\sqrt{ \lambda' } }  \right) ,
%\label{eq:partialtA}
%\\
%%\end{align}
%%\begin{align}
%\partial_z A_\pm (t, z) &= \Theta(z) \frac{e^{\pm i \m z}}{2} \left[ - j_\pm(t - z)
%%\right.
%%\nonumber\\ &~~~~
%%\left.
%+ \int^{t - z}_{-\infty} \! dt_0 j_\pm(t_0) \left( \pm i \m I_0 (\sqrt{\lambda'}) - \frac{\m^2 z I_1 (\sqrt{\lambda'}) }{ \sqrt{\lambda'}} \right) \right],
%\label{eq:partialzA}
%\end{align}
%%\end{widetext}
%where in analogy with \eqref{eq:lambda} we define $\lambda' \equiv \m^2( (t-t_0)^2 - z^2)$. The familiar behavior is recovered in the Standard Model limit $\mu \rightarrow 0$, with
%\begin{align}
%\partial_t A^{\mu=0}_\pm(t, z) &= \Theta(z) \frac{1 }{2} j_\pm(t-z) , \\
%\partial_z A^{\mu=0}_\pm(t, z) &= -\Theta(z) \frac{1 }{2} j_\pm(t-z) .
%\end{align}

\subsection{Green Function Solution in Four Dimensions} \label{sec:green4d}

The Green function \eqref{eq:greensoln2d} is valid for sources that are spatially uniform in the $x$ and $y$ directions. In many situations, including laser pulses and the light from distant stars, this approximation is  sufficient. However, in other cases the fully four-dimensional Green function may be relevant.
As we show in this section, most of the 4D solution can be written in terms of the 2D Green function derived in Section~\ref{sec:green2d}. One of the new integrals cannot be so easily solved analytically, but with some effort it can be put in the form of a rapidly converging infinite series for easier numerical evaluation.

In the Lorenz gauge with neutral sources, $\nabla \cdot \vec{A} = 0$, $A^0 = 0$, the Green function for $\vec{A}$ satisfies % \bl{For transverse sources, $\rho = 0 \longrightarrow A^0$, and the Lorentz and Coulomb gauges reduce to the same eom (below)}
\begin{align}
%\left( \delta_{ij} \partial^2 - \frac{\alpha \dot\theta}{\pi} \epsilon_{\ell i k} \partial_\ell \right) G_{k j}(x^\mu - y^\mu) = \delta_{ij} \delta^{(4)}(x^\mu - y^\mu),
\left( \delta_{ij} \Box - \dot\theta  \epsilon_{\ell i k} \nabla_\ell \right) G_{k j}(x^\mu - y^\mu) = \delta_{ij} \delta^{(4)}(x^\mu - y^\mu),
\end{align}
where the cross product term in \eqref{eq:AEDeom} forces the Green function $G_{kj}$  to have a nontrivial tensor structure. Its Fourier transform $\tilde{G}_{kj}$ satisfies
\begin{equation}
\left[ (-\omega^2 + k^2) \delta_{ik} + 2i \m k_\ell \epsilon_{\ell i k } \right] \tilde{G}_{kj}(\omega, \vec{k}) = \delta_{ij} \, ,
\end{equation}
in terms of $\m$ from \eqref{eq:em}, frequency $\omega$, and $k^2 = k_i k^i$ for $i=1,2,3$.
Inverting the operator that acts on $\tilde{G}_{kj}$, the Green function can be written as
\begin{align}
\tilde{G}_{kj} = \tilde{A} \delta_{kj} + \tilde{B} k_\ell \epsilon_{\ell kj}  + \tilde{C} k_k k_j \, ,
\end{align}
where
\begin{align}
\tilde{A} = - \frac{\omega^2 - k^2}{\beta},
&&
\tilde{B} = - \frac{2i \m}{\beta} ,
&&
\tilde{C} = \frac{4\m^2}{(\omega^2 - k^2 ) \beta },
&&
\beta = (\omega^2 - k^2)^2 - 4 k^2 \m^2,
\end{align}
where the four roots of $\beta$ are
\begin{equation}
\omega_\pm^2 = k^2 \pm 2k \m.
\end{equation}
Since the current sources are transverse, the Green function can be simplified by the transverse projection
\begin{align}
\tilde{G}^T_{ij} &\equiv \left( \delta_{ik} - \frac{k_i k_k}{k^2 } \right) \tilde{G}_{kj}\nonumber\\
&= \tilde{A} \left( \delta_{ij} - \frac{k_i k_j}{k^2 } \right) + \tilde{B} k_\ell \epsilon_{\ell i j }, \\
G^T_{ij} &= \delta_{ij} A + i \epsilon_{\ell i j } \nabla_\ell B + A_{ij},
\end{align}
where $A$, $B$ and $A_{ij}$ are the Fourier transforms of $\tilde{A}$, $\tilde{B}$, and $-\tilde{A} k_i k_j/k^2$, respectively.

Both $A$ and $B$ can be written in terms of the scalar function $G_0(t, z)$ from the two-dimensional case, \eqref{eq:greensoln2d},
\begin{align}
%A &= - \frac{1}{2} \sum_{\pm} \int\! \frac{d\omega\, d^3 \vec{k} }{(2\pi)^4} \frac{e^{i (\omega t - \vec{k} \cdot \vec{x} ) }  }{\omega^2 \pm 2k \m - k^2 } \\
%&= \frac{1}{4 \pi r} \partial_r \int_{- \infty}^\infty \! \frac{dk}{2\pi} \frac{d\omega}{2\pi} \left( \frac{ \cos(k r) e^{i \omega t}  }{\omega^2 - (k - \m)^2 + \m^2 } +  \frac{ \cos(k r) e^{i \omega t}  }{\omega^2 - (k + \m)^2 + \m^2 } \right) \\
%&= \left.\left.\frac{1}{4 \pi r} \partial_r \right( \cos(r \m ) G_0( t, r; - \m^2)  + \cos(r \m ) G_0(t, -r; -\m^2) \right),
A &= \left.\left.\frac{1}{2 \pi r} \partial_r \right( \cos(r \m ) G_0( t, r; - \m^2)   \right), \nonumber\\
%B = \left.\left. \frac{\sin(r \m) }{4 \pi r} \right( G_0( t, r; - \m^2)  + G_0(t, -r; -\m^2)  \right),
B&= \left.\left. \frac{\sin(r \m) }{2 \pi r} \right( G_0( t, r; - \m^2)    \right),
\label{eq:ABsoln}
\end{align}
where
\begin{equation}
G_0 (t, r; -m^2) \equiv  \frac{1}{2} \Theta(t)  \Theta(t^2 - r^2) I_0(\sqrt{ m^2 t^2 - m^2 r^2}).
\end{equation}
%Similarly, $B$ can be written as
%\begin{align}
%%B &= \frac{1}{2} \sum_\pm \frac{\pm1}{\pi r} \int_0^\infty \! \frac{dk}{2\pi} \sin(k r) \int_{-\infty}^\infty \! \frac{d\omega}{2\pi} \frac{i e^{i\omega t} }{\omega^2 \pm 2 k \m - k^2 } \\
%%&= \frac{i}{4 \pi r} \int_{- \infty}^\infty \! \frac{dk}{2\pi } \frac{d\omega}{2\pi} \left( \frac{\sin(kr) e^{i \omega} }{\omega^2 - (k - \m)^2 + \m^2 } -  \frac{\sin(kr) e^{i \omega} }{\omega^2 - (k + \m)^2 + \m^2 } \right) \\
%B &= \left.\left. \frac{\sin(r \m) }{4 \pi r} \right( G_0( t, r; - \m^2)  + G_0(t, -r; -\m^2)  \right).
%\end{align}
Attempting the same technique for $A_{ij}$ leads to the incomplete expression
\begin{align}
%A_{ij} &= -\frac{1}{2} \partial_i \partial_j \sum_\pm \int\! \frac{d \omega\, d^3 \vec{k} }{(2\pi)^4 } \frac{1}{k^2} \frac{e^{- (\omega t - \vec{k} \cdot \vec{x} ) }  }{\omega^2 \pm 2k \m - k^2}  \\
%&= - \frac{1}{4} \partial_i \partial_j \sum_{\pm} \int_{-1}^1 \! \frac{d \cos\theta}{2\pi} \int_{-\infty}^\infty \frac{dk}{2\pi} \frac{d\omega}{2\pi} \frac{e^{i (\omega t - k r \cos\theta) }}{\omega^2 \pm 2 k \m - k^2 } \\
A_{ij}  &=  \frac{1}{4\pi} \partial_i \partial_j \int_{-1}^1 \! dq\, \cos(q r \m ) G_0( t, qr; -\m^2) ,
\label{eq:intAij}
\end{align}
an integral that does not have a simple expression in terms of Bessel functions or other hypergeometric functions.
In Appendix~\ref{appx:integral} we show how the integral form of $A_{ij}$ can be replaced with an infinite series over a product of hypergeometric functions, with the result 
\begin{align}
A_{ij} &= \frac{1}{8 \pi} \partial_i \partial_j \left[ \Theta(t) \Theta(t^2 - r^2) \sum_{\ell=0}^\infty \frac{\left(- \frac{1}{4} \m^2 r^2 \right)^\ell}{\ell! (\ell+\frac{1}{2})} 
 \left(\frac{2}{|\m t |}\right)^\ell I_\ell(| \m t |)
\ {_1 F_2}\! \left.\left( \begin{array}{c} \ell+\frac{1}{2} \\ \frac{1}{2}, \frac{\ell}{2} + \frac{1}{4}  \end{array} \right| - \frac{1}{4} \m^2 r^2 \right)
\right].
\label{eq:Aijsoln}
\end{align}
The series in $\ell$ converges rapidly %due to the $(\ell !)$ denominator factor and the fact that the series expansion of$_1F_2$ function itself converges 
for $\mu r \leq 2$. For large $\mu r \gg 1$ and $\mu t \gg 1$ it converges for $\ell > \ell_\text{max}$, for an $\ell_\text{max}(\mu t, \mu r)$ given in Appendix~\ref{appx:integral}.

\subsection{Application to Monochromatic Signals} \label{sec:constnumerics}

The Green function for the vector potential $\vec{A}$ (Eq.~(\ref{eq:greensoln2d})) and its derivatives exhibit novel inside-the-lightcone components, which  induce exponentially growing, semi-static residual fields in the wake of a signal.
%The modified phase velocity also 
The on-the-lightcone contribution to the signal is modified as well, as a result of the perturbed phase velocity. 
% in addition to the modified on-the-lightcone contribution to the propagating signal associated with the changing phase velocity.
Both of these effects provide signatures of the axion background in the path of an electromagnetic wave.
In this section, we provide a few examples to show how simple monochromatic signals can be distorted %in the $\dot\theta \approx \textit{constant}$ fuzzy dark matter limit, $m_a \ll \dot\theta$.
on timescales $T$ that are shorter than the period of axion oscillation, $T \ll m_a^{-1}$.

\subsubsection{Phase Velocity}

One distinctive feature of the modified electrodynamics, the helicity-dependent phase velocities, can be quantified directly from the equations of motion as in Refs.~\cite{DeRocco:2018jwe,Obata:2018vvr}. From the dispersion relations for right- and left-polarized light, \eqref{eq:dispersion}, the phase velocities can be expanded in powers of $\m = \frac{1}{2} \dot\theta(t)$,
\begin{align}
v_\text{phase} = \frac{\omega_\pm}{k} %= \frac{\sqrt{k^2 \pm 2 k \m} }{k} 
= \sqrt{ 1 \pm 2 \frac{\m}{k} } 
= 1 \pm \frac{\m}{k} - \frac{1}{2} \frac{\m^2}{k^2} + \mathcal O(\m^3/k^3).
\label{eq:vphase}
\end{align}
This result can be also be recovered from the Green function in \eqref{eq:greensoln2d}. If the signal is driven by a monochromatic source, 
\begin{equation}
j_\pm(\tau) = e^{i \Omega \tau} f_\pm (\tau),
\end{equation}
the solutions for $A_\pm(t, z)$ are proportional to trigonometric factors of $\exp(i \Omega (t - z) \pm \m z )$, reproducing the linear order term in \eqref{eq:vphase}.
%In Appendix~\ref{sec:monochromatic} we provide the explicit solutions for the $\vec{E}$ and $\vec{B}$ fields for this simple example.

Extremely sensitive measurements of the phase shift induced by gravitational waves are the bedrock for the remarkable recent detections of black hole and neutron star mergers. The sensitivity of Advanced LIGO~\cite{Martynov:2016fzi} to the gravitational strain $h = \Delta L/L$ exceeds $10^{-23}/\sqrt{\text{Hz}}$ for gravitational waves with frequency $f \sim 10^2\, \text{Hz}$. An axion interferometer comparing the phases of left- and right- polarized laser beams of angular frequency $\Omega$ observes a phase difference 
\begin{align}
\Delta \phi \equiv \phi_+ - \phi_- = 2 \m L = \dot\theta(t) L
\end{align}
 after the laser propagates a length $L$. Compared to the equivalent phase shift corresponding to a change in the path length $\Delta L$, $\Delta \phi = \Omega \Delta L$, 
the technology capable of detecting an $h_\text{min} \sim 10^{-23}$ would also be able to set a limit on $\m/\Omega$ of order
\begin{equation}
\frac{2 \m }{\Omega } \lesssim h_\text{min}\, ,
\label{eq:hmin}
\end{equation}
in the context of an axion interferometer.
To use the 1064~nm laser of LIGO~\cite{Abbott:2007kv} as an example  ($\Omega = 1.77\times 10^{15}\, \text{rad/s}$),
an interferometer capable of similar precision has a potential sensitivity to any $\dot\theta \gtrsim 2 \times 10^{-8}\, \text{Hz}$, which overlaps with the parameter space indicated by \eqref{eq:dottheta} for $10^{-11}\, \GeV^{-1} \lesssim g_{a\gamma\gamma} \lesssim 10^{-10}\, \GeV^{-1}$.

%In fact, for $\mu \ll \Omega$, the $m_a \rightarrow 0$ limit can be taken without insisting that $m_a \ll \dot\theta$ as long as the characteristic length $L$ and integration time $T$ are both small compared to $m_a^{-1}$.
To achieve the sensitivity indicated by \eqref{eq:hmin}, the approximation that $\dot\theta(t) = 2\m$ is constant needs to hold only for as long as it takes the photon beam to traverse the interferometer. 
% This requirement implies that a detector of characteristic length $L$ is sensitive only to axions of mass $m_a < L^{-1}$. % In this case the Compton wavelength of the axion is also large enough compared to $L$ to justify neglecting the spatial gradients $\nabla \theta$. Otherwise, as the axion oscillates its contributions to the phase velocities in \eqref{eq:vphase} tend to cancel out. 
As long as $m_a L \ll 1$ and $m_a T \ll 1$ for the characteristic length $L$ and time-of-flight $T$ for the measurement, it is not necessary to insist that $m_a \ll \dot\theta$.
 A detector with $c T \sim n L \sim \mathcal O(10^3\, \text{km})$ (where $n$ is the effective number of reflections of the light within the chamber) would thus be sensitive to $m_a \lesssim 10^{-12}\, \eV$, 
% where $n$ indicates the 
while more massive axions would be more easily detected with shorter interferometers.
In the case of LIGO, the beam cavity storage time $T$ is long compared to the length of each arm, with $T \sim n L/c$ for some $n \approx 70$~\cite{Abbott:2007kv}.

% {\bf\color{blue} PD: I  think we need a more explicit caveat here that LIGO achieves its sensitivity for $L\sim 1$ km and $n\gg 1$, if that is indeed true (I thought $n$ was several hundred.) We want to keep the sensitivity fixed so that the sensitivity to $g_{a\gamma\gamma}$ doesn't get much worse. Maybe that means we should quote also $T \sim 10^3 L \sim \mathcal O(1000\, \text{km})$ is sensitive to $m_a \lesssim 10^{-12}\, \eV$. The scaling is obvious but this is just the sort of thing a referee would enjoy getting irritable about.}
%An 

\subsubsection{After-Pulse Residual Fields}

A substantively new effect appears when we consider the Green function to all orders in $\dot\theta$, for signals of finite duration $T$. In the $\dot\theta = 0$ vacuum of standard electrodynamics, such a signal propagates away from the source at speed $c$, maintaining the same duration $T$.
In the $\dot\theta \neq 0$ background, this is no longer true. Rather than returning to zero after the signal passes, the $\vec{E}$ and $\vec{B}$ fields retain a residual nonzero value that grows with time. 
Interpreting $-\frac{1}{4} \dot\theta^2$ as a tachyonic mass term for the photon, the growth of the $\vec{E}$ and $\vec{B}$ fields is a consequence of the tachyonic instability---a transfer of energy from the axion background into long-wavelength photons triggered by the original signal. This growth takes place inside the lightcone, rather than strictly on it.

Again specializing to a monochromatic signal $j_\pm(\tau) = 2 e^{i \Omega \tau} f_\pm(\tau)$ for simplicity, and taking the source at $z=0$ to satisfy $j_\pm(\tau) = 0$ for $\tau <0$ and $\tau > T$,  the ``residual'' field $A^{(r)}_\pm$ at $z >0$ refers to the nonzero field value after $t - z > T$. We have 
\begin{align}
\partial_t A^{(r)}_\pm(t- z > T) &= \m \int_{0}^{T} \! d\tau\, e^{i \Omega \tau\pm i \m z} f_\pm(\tau) \frac{(\m t - \m \tau) I_1( \sqrt{\lambda'} ) }{\sqrt{ \lambda' } }  ,
\label{eq:residualAt}
\\
\partial_z A^{(r)}_\pm(t- z > T)  &=  \m \int^{T}_{0} \! d\tau e^{i \Omega \tau\pm i \m z} f_\pm(\tau)  \left( \pm i  I_0 (\sqrt{\lambda'}) - \frac{\m z I_1 (\sqrt{\lambda'}) }{ \sqrt{\lambda'}} \right) ,
\label{eq:residualAz}
\end{align}
where $\lambda' = \m^2 (t- \tau)^2 - \m^2 z^2 $.
In the high frequency limit, the rapid oscillations of $e^{i \Omega \tau}$ tend to cancel out the contributions from both integrals, so that the strengths of the residual $\vec{E}$ and $\vec{B}$ fields are proportional to $\m/\Omega$.
However, especially for $\mu T \sim \mathcal O(1)$, the residual fields can become substantial: for $\lambda' \gtrsim \mathcal O(1)$, the exponential growth of the Bessel functions becomes apparent, and eventually compensates for the $\mu/\Omega$ suppression.

\begin{figure}[t]
\centering
\includegraphics[height=0.52\textwidth]{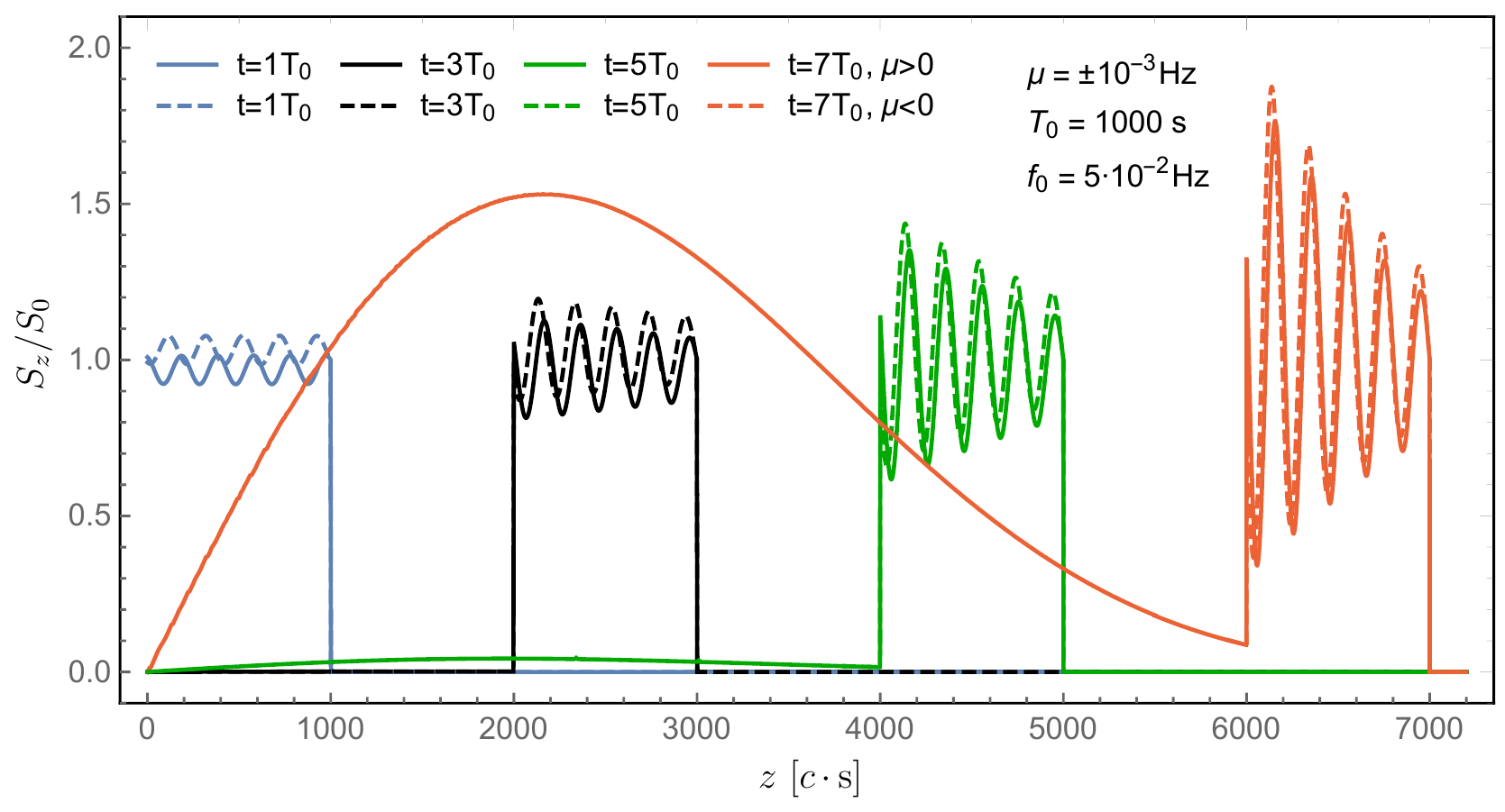}
\caption{The spatial profile the Poynting vector $S_z(z)$ of a propagating right-polarized square pulse  is shown as a function of $z$, at several snapshots in time: $t = \{T_0, 3T_0, 5T_0, 7 T_0\}$ with $T_0 = 10^{3}\, \text{s}$, where the solid and dashed lines correspond respectively to $\m = \pm 10^{-3} \, \text{Hz}$. For $0 \leq t-z \leq T_0$ (``on the lightcone''), the difference in sign affects the phase velocities and spectrum of the pulses, but for $t-z > T_0$ (``inside the lightcone'') the sign of $\m$ is irrelevant for $S_z(t,z)$.}
\label{fig:propagation}
\end{figure}

To demonstrate the distortion to a signal as it propagates through space, Figure~\ref{fig:propagation} shows the Poynting vector $\vec{S} = \vec{E} \times \vec{B}$
as a function of $z$ at four snapshots in time. At $t = T_0$, when the signal is newly produced, it exhibits relatively mild modifications to the original $S_z(t,z) = S_0$ square wave, in this example with $T_0 = \m^{-1}$.
By the $t = 5 T_0$ snapshot, not only has the signal become notably distorted, but it also has developed nonzero values inside the lightcone, of magnitude $S_z(t - z > T_0) \sim 10^{-1} S_0$. At $t = 7 T_0$ this part of the field exceeds $S_z(t - z > T_0) \gtrsim S_0$, and continues to grow exponentially for $t > 7T_0$.

\begin{figure}[t]
\centering
\includegraphics[height=0.48\textwidth]{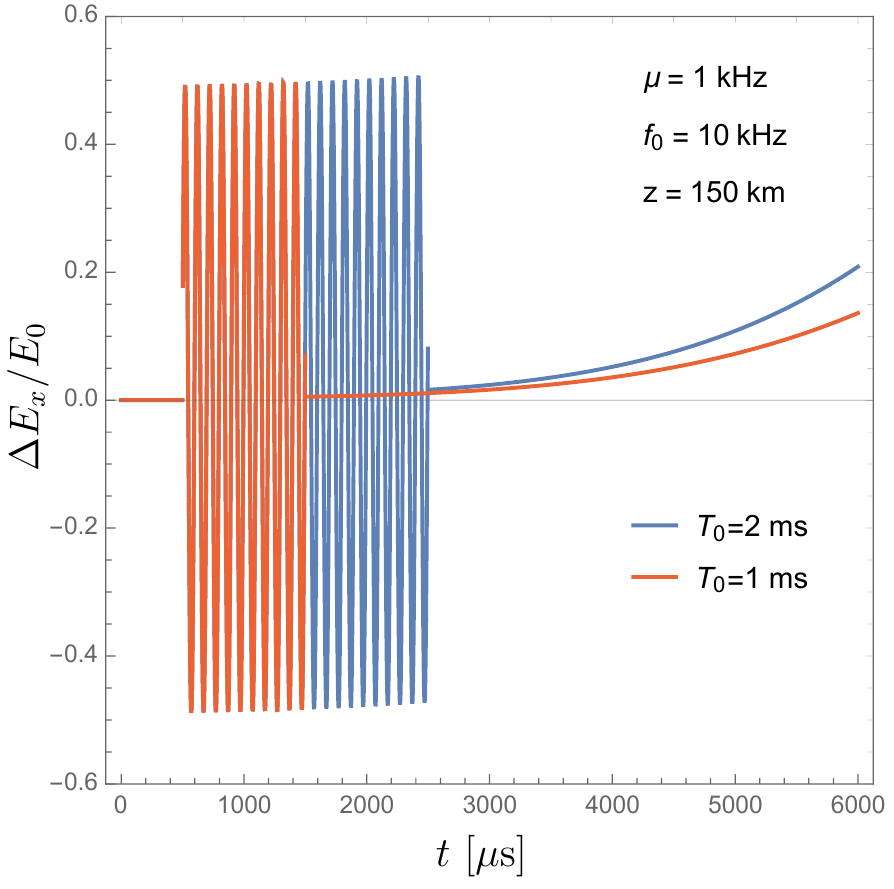}
\includegraphics[height=0.48\textwidth]{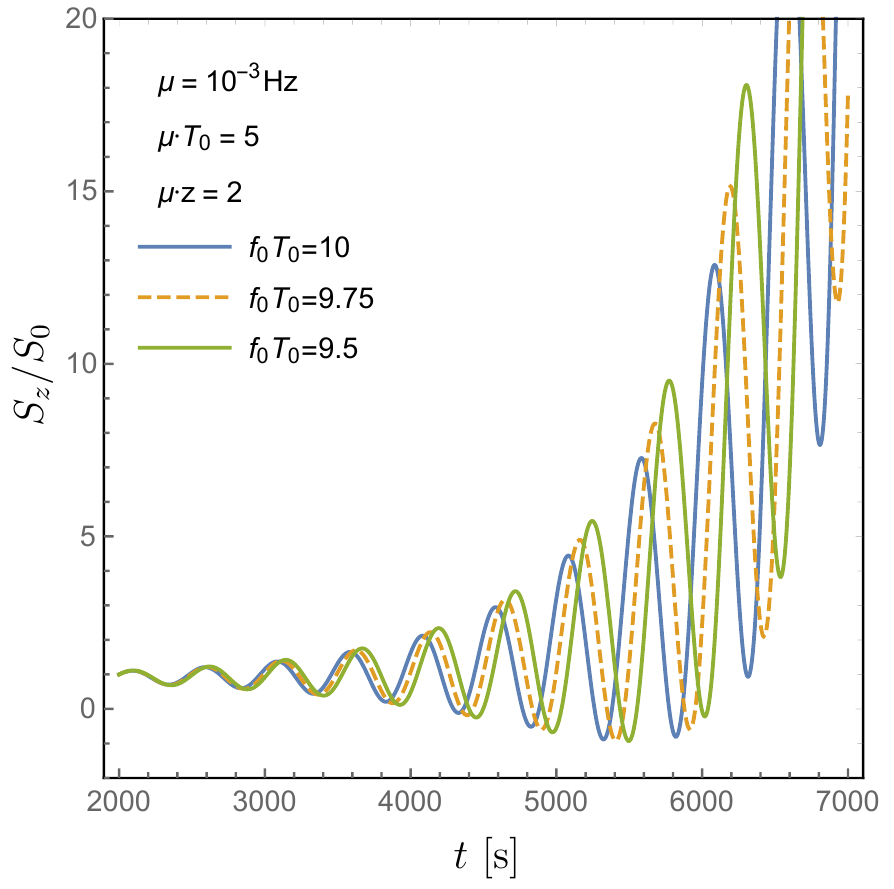}
\caption{Left: the $E_x$ field is shown as a function of time at fixed $z = 150\, \text{km}$ for square pulses of frequency $f_0 = \Omega/2\pi =10\,\text{kHz}$, and durations $1000\,\text{$\mu$s}$ (orange) and $2000\,\text{$\mu$s}$ (blue). Oscillations begin as the front of the pulse passes through the fixed value of $z$, followed at later times by monotonic growth. For illustration we use an inflated value of $\m = 1\,\text{kHz}$. Not pictured, the $\vec{B}$ and $E_y$ fields have a nearly identical profile. 
Right: The value of the Poynting vector $S_z$ is shown as a function of time for three square pulses of duration $T_0 = 5\cdot10^{3}\,\text{s}$ and similar frequencies $f_0 \sim 10/T_0$, measured at a distance $z = 2/\m$ from the source, for $\m = 10^{-3}\, \text{Hz}$. 
In the $\m = 0$ case the Poynting vector would be constant, with $S_z^{m=0}(0 \leq t-z \leq T_0) = S_0$.
}
\label{fig:residual}
\end{figure}

In Figure~\ref{fig:residual}, we show $E_x(t)$  for two square pulses with different durations $T$,
to demonstrate the relationship between the pulse length and the strength of the residual fields.
In this example we inflate the typical value of $\mu$ to $\mu = 1\,\text{kHz}$ so that the strength of the residual pulse approaches the amplitude of the original signal for $t \sim \mathcal O(\text{few}) \times T$ in an example with $10\,\text{kHz}$ radiation. However, the same plots can be reused for any value of $\mu$ by rescaling $\Omega$, $t$ and $z$ so as to keep $\mu \Omega$, $\mu t$ and $\m z$ constant, as in the right hand panel of Figure~\ref{fig:residual}.
For another example, using $\mu = 10^{-8}\, \text{Hz}$ instead of $10^{3}\, \text{Hz}$, the plot in Figure~\ref{fig:residual} would show $t$ in units of $10^5\, \text{s}$ rather than $\mu\text{s}$, and $z = 1.5\times 10^{13}\, \text{km} = 0.49\, \text{pc}$.

\subsubsection{Exponential Growth} \label{sec:exponential}

At sufficiently late times, $\m t \gtrsim 1$, the growth in the $\vec{E}$ and $\vec{B}$ fields highlighted in Figure~\ref{fig:residual} becomes exponential, driven by the Bessel functions $I_0(\sqrt{\lambda})$ and $\lambda^{-1/2} I_1(\sqrt{\lambda})$ in the Green functions Eqs.~(\ref{eq:residualAt}--\ref{eq:residualAz}).
This is true not only of the semi-static residual fields in the wake of the signal, but also for the signal itself, if its duration $T$ is long compared to $\m^{-1}$.
For the phenomenologically relevant values of $\mu$ %at the present time, the $\dot\theta \lesssim 5 \cdot 10^{-7}\, \text{Hz} \times \sqrt{\rho_\chi} (0.15\,\eV)^{-2}$ of 
given by \eqref{eq:dottheta}, probing the strongly exponential behavior requires extremely long coherence times $T_0 >\m^{-1} \gtrsim 10^{8}\,\text{s}$ for both the axion field and the radiation source, unless the axion density is significantly enhanced beyond the expected $\rho_a \sim 0.4\, \GeV/\text{cm}^3$.
Nevertheless, it is a useful exercise to explore the behavior of the fields in this extreme limit.

%\begin{figure}[t]
%\centering
%\includegraphics[height=0.48\textwidth]{plots/growth-exponential-sZ.pdf}
%\includegraphics[height=0.48\textwidth]{plots/growth-exponential-sZ-after-alt.pdf}
%\caption{The value of the Poynting vector $S_z$ as a function of time is shown for square pulses of duration $T_0 = 5\cdot10^{3}\,\text{s}$, with similar frequencies $f_0 \sim 10/T_0$, measured at a distance $z = 2/\m$ from the source, for $\m = 10^{-3}\, \text{Hz}$. The left panel shows $S_z(t,z)$ with $0 \leq t-z \leq T_0$, which in the $\m = 0$ case would be constant, $S_z^{m=0}(t, z) = S_0$. On the right, we show the exponential growth in $S_z(t, z)$ for $t > z + T_0$, where standard electrodynamics predicts $\vec{E} = \vec{B} = 0$. In this example with $\m z =2$ and $\m T_0 = 5$, the radiated power is not very sensitive to the phase $\Omega T_0$, as can be seen from the close spacing of the lines in the right-hand plot.
%}
%\label{fig:exponentialgrowth}
%\end{figure}

Even in the late-time limit $\m (t-z) >1$, there is a clear distinction between the received ``signal'', when $0 \leq t-z \leq T$, and the ``residual'' fields, $t-z > T$. In contrast to the relatively mild modifications to the signal in Figure~\ref{fig:residual}, the signal eventually becomes significantly amplified and distorted for large $\m t \gg1$ and $\m z\gg1$.
The semi-static residual fields that appear in the wake of the signal continue to grow exponentially, until the point where the energy density in the electromagnetic fields becomes comparable to $\rho_a$, and the back-reaction on the axion field can no longer be neglected.

Figure~\ref{fig:residual} (right panel) shows the modified signal for an example with $\m = 10^{-3}\, \text{Hz}$, where the pulse duration ($T = 5\times 10^3\,\text{s}$) and propagation distance ($z=6\times 10^8\, \text{km}$) are both larger than $\m^{-1}$.
In the $\m\rightarrow 0$ limit of standard electrodynamics, the power density $\abs{S_z}$ of this circularly-polarized square wave would remain constant, $S_z^{m = 0} (0 \leq t-z \leq T) = S_0$, returning to zero for $t-z > T$.
Instead, the received power in the $\m \neq 0$ case varies as a function of time, oscillating with ever-larger fluctuations and increasing exponentially. Once $\m(t - z) \gtrsim 1$, the fluctuations in the power $\Delta S_z$ become larger than the magnitude of the power at the source, $S_0$, and the exponential growth soon ensures that $S_z(t) \gg S_0$ for all $t-z \gg \m^{-1}$.

After the signal has passed by, the residual fields are well described by Eqs.~(\ref{eq:residualAt}--\ref{eq:residualAz}), which in the $\lambda' \gg 1$ limit closely resemble the asymptotic expansion described in \eqref{eq:besselasymptotic} in Appendix~\ref{appx:integral}.
The dominant term in this expansion is the exponential
\begin{equation}
\lim_{\lambda \rightarrow \infty} I_j(\sqrt{\lambda}) \approx \frac{e^{\sqrt\lambda} }{\sqrt{2\pi \sqrt{\lambda}}}.
\end{equation}
%Despite the oscillatory integrands in Eqs.~(\ref{eq:residualAt}--\ref{eq:residualAz}), the residual power is not very dependent on the phase $e^{i\Omega T_0}$, as one would expect if the amplitude of the integrand varied smoothly. %Instead, the exponential growth ensures that only the largest values of $\lambda' = \m^2(t-\tau)^2 - \mu^2 z^2$ are relevant for the integral
%Instead, due to the exponential growth of the Bessel functions, the precessing phase $e^{i \Omega \tau}$ is only efficient at cancelling the integrands in the high frequency limit, $\Omega \gg \m$.

\begin{figure}[t]
\centering
\includegraphics[width=0.48\textwidth]{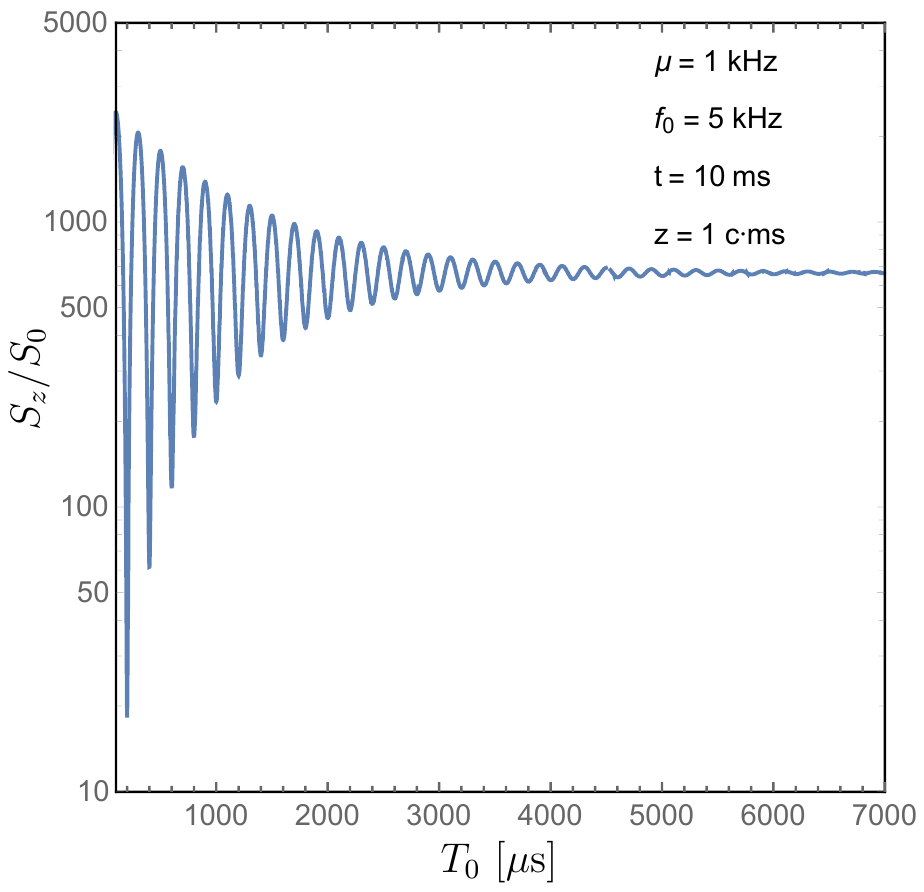}
\includegraphics[width=0.48\textwidth]{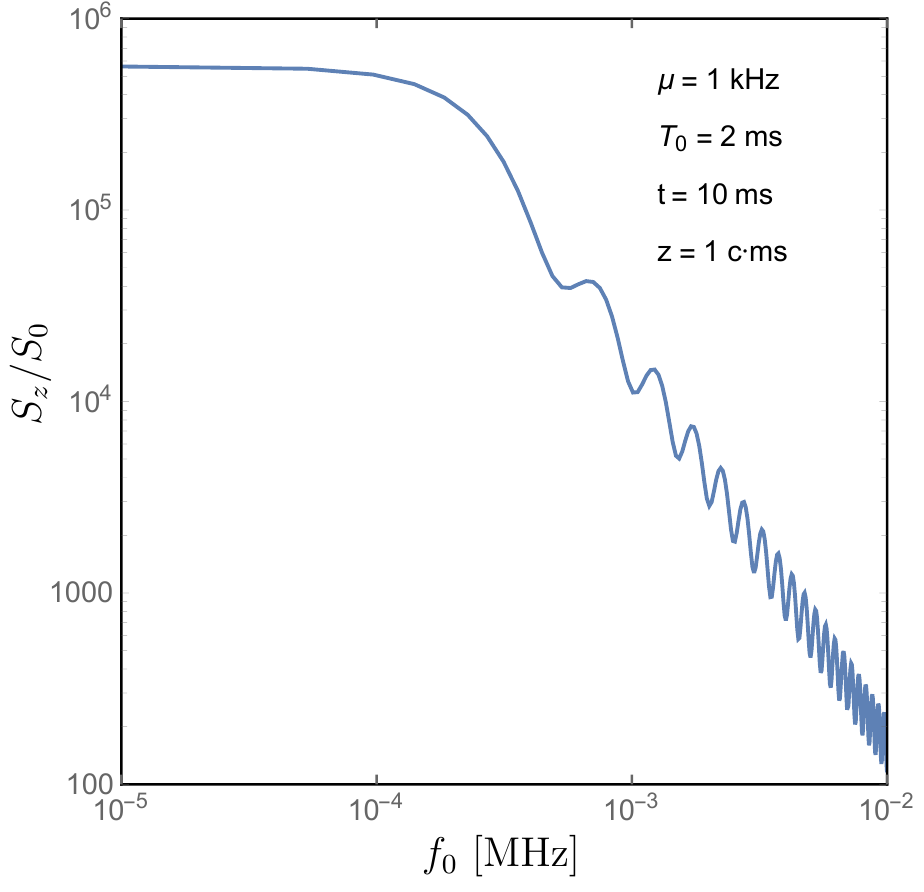}
\caption{Left: the fractional change in the radiation power $S_z$ at $z = 300\, \text{km}$, for $\mu = 1\, \text{kHz}$, as a function of pulse duration $T_0$, measured at $t = T_0 + 10\,\text{ms}$. The approach to a constant value as the pulse duration increases reflects the fact that it is the  first part of the signal which has induced the largest contribution to the growing instability.
Right: $\Delta S_z/S_0$ as a function of radiation frequency $f = \Omega/2\pi$, for the same values of $\m$ and $z$, for fixed $t = 10\, \text{ms}$ and $T_0 = 2\,\text{ms}$. The falling value of $\Delta S_z$ mimics the $\Delta S_z \propto \Omega^{-2}$ scaling indicated in Section~\ref{sec:exponential}.
The oscillations in the received power for $t-z > T_0$ correspond to integer values of $\pi \Omega T_0$. 
As shown in the left panel, small variations in $\Omega T_0$ have a substantial impact on the power when $\m T_0 \lesssim \mathcal O(\text{few})$, but for larger values of $T_0$ %($\m T_0 \gtrsim \mathcal O(\text{few})$) 
the contribution to $S_z(t, z)$ from the latter part of the signal is dwarfed by the
 exponential growth seeded by the first part.
}
\label{fig:timingfrequency}
\end{figure}

To show the dependence of the strength of the residual fields on the pulse length $\Omega T_0$, the left panel of Figure~\ref{fig:timingfrequency} shows $S_z(t, z)$ at $t = 10\, \text{ms}$, $z = 1\, \text{ms}$, as a function of pulse duration $T_0$, with fixed frequency $f_0= \Omega/2\pi = 5 \,\text{kHz}$.
Once $T_0 \gtrsim 4 \mu^{-1}$, the strength of the field at $(t, z)$ approaches a constant value. 
Evidently, $S_z (t-z > T)$ is driven primarily by the first part of the signal, which has induced the longest-lived instabilities: the subsequent exponential growth of the initial instability makes this part of the signal the most consequential.

By considering the high-frequency limit of Eqs.~(\ref{eq:residualAt}--\ref{eq:residualAz}), where $f_\pm(\tau)$ and $I_j(\sqrt{\lambda'})$ both vary slowly compared to $e^{i \Omega \tau}$, it is easy to see that the inside-the-lightcone electromagnetic fields scale as $|\vec{E}| \propto \Omega^{-1}$ and $|\vec{B}|\propto \Omega^{-1}$,
so that the radiation density is proportional to $S_z \propto \Omega^{-2}$.
In the right panel of Figure~\ref{fig:timingfrequency}, the power $S_z(t, z)$ is shown for the same values of $t$ and $z$, this time as a function of the frequency of the square pulse ($f_0 = \Omega/2\pi$), with fixed pulse duration $T_0 = 2\cdot 10^{-3}\, \text{s}$.
At low frequencies, $f_0 \ll T_0^{-1}$, the power approaches a constant. In this limit the phases of the integrands in Eqs.~(\ref{eq:residualAt}--\ref{eq:residualAz}) are essentially independent of $\Omega$, $e^{i \Omega \tau \pm i \m z} \approx e^{\pm i \m z}$.
In the opposite limit, $f_0 \gg T_0^{-1}$, the $S_z \propto \m^2/\Omega^2$ scaling suggested by  Eqs.~(\ref{eq:residualAt}--\ref{eq:residualAz}) becomes manifest.
Together, the $T_0 \gg \m^{-1}$ and $f_0 \gg \m$ limits indicate that the power scales roughly as $S_z(t, z) \sim e^{\sqrt{\lambda}} (\m/\Omega)^2$, for $\lambda = \m^2 t^2 - \m^2 z^2$ with $t-z \gg T_0$.
In many applications involving radio, visible or X-ray radiation, the hierarchy between $\mu$ and $\Omega$ imposes an extreme penalty on the magnitude of the surplus power, so that only after multiple $e$-foldings would it be possible to detect the signal.

\medskip

Our derivation of the Green function \eqref{eq:greensoln2d} assumes that $\ddot \theta$ vanishes, so that $\mu(t)$ can be treated as constant.
For a standard axion oscillating in a quadratic potential, $\mu(t)$ varies according to
\begin{align}
\mu(t) = \mu_0 \cos(m_a t), 
&&
\mu_0 = \frac{1}{2} g_{a\gamma\gamma} \sqrt{\rho_a}.
\label{eq:mu0cost}
\end{align}
At times $T$ comparable to the period of axion oscillation, $T \sim m_a^{-1}$, the steady-state approximation $\mu(t) \simeq \mu$ is no longer valid, and it is necessary to use the methods described in Section~\ref{sec:osc}.
In principle, because $\mu_0$ is set directly by $\rho_a$ and $g_{a\gamma\gamma}$, it is independent of $m_a$, and the exponential growth in the late-time limit $\mu T\gg 1$ can be explored without violating the steady-state condition $m_a T \ll 1$.
Achieving $\mu_0 \gg m_a$ does, however, require a non-standard ALP model with an extended field range. \eqref{eq:mu0cost} implies
\begin{align}
\mu(t) = \mu_0 \cos(m_a t) \longrightarrow \theta(t) = \theta_0 \sin(m_a t), && \theta_0 = 2 \frac{\mu_0}{m_a},
\end{align}
so $\mu_0 \gg m_a$ necessarily implies that $\theta_0 \gg 2$.

In practice, this region of parameter space with $\mu_0 > m_a$ is mostly ruled out by experiment. %, as we describe in Section~\ref{sec:osc}.
Based on the constraints for $g_{a\gamma\gamma}$~\cite{Dicus:1979ch,Raffelt:2006cw, Anastassopoulos:2017ftl,Brockway:1996yr,Grifols:1996id,Conlon:2017qcw} and $m_a$~\cite{Armengaud:2017nkf,Irsic:2017yje,Kobayashi:2017jcf,Nori:2018pka},
for dark matter densities of $\rho_a \approx 0.4\, \GeV/\text{cm}^3$,
\eqref{eq:dottheta} indicates that $\dot\theta \ll m_a$ for all allowed $m_a$ and $g_{a\gamma\gamma}$. 
To realize the late-time exponential growth within the steady-state axion background, three conditions must be satisfied:
$m_a$ must be near the low-mass ``fuzzy dark matter'' extreme, $m_a \sim 10^{-21}\, \eV$;
the coupling $g_{a \gamma\gamma}$ must be relatively strong, $g_{a\gamma\gamma} \lesssim 10^{-11}\, \eV$; 
and the value of $\sqrt{\rho_a}$ in the path of the photon must be enhanced by a few orders of magnitude, for example by concentrating some fraction of the axions  into  dense clumps or ``axion stars.''
If these conditions are not satisfied, then $m_a T \ll 1$ implies $\mu_0 T \ll 1$, 
%and the oscillation of the axion field must be taken into account before we can investigate the exponential growth. 
and the Green function is well approximated by its series expansion in $\lambda$.

\section{Oscillating Axion Background} \label{sec:osc}

In an oscillating axion background,
\begin{align}
\dot\theta(t) = \dot\theta_0 \cos(m_a t),
&&
\dot\theta_0 = g_{a\gamma\gamma} \sqrt{\rho_a},
\end{align}
 the Green function for the vector potential $A_\pm$ in $(1+1)$ dimensional spacetime satisfies
\begin{equation}
(\partial_t^2 - \partial_z^2 \pm i \dot\theta_0 \cos(m_a t) \partial_z)  g_\pm(t - t_0, z - z_0) = \delta(z - z_0) \delta(t - t_0).
\label{eq:greenDEosc}
\end{equation} 
As in Section~\ref{sec:green2d} we restrict our analysis to propagating plane waves.
Unlike the $\dot\theta \approx \textit{const.}$ limit of \eqref{eq:G0int}, \eqref{eq:greenDEosc} cannot be inverted to find an algebraic expression for the Fourier transform of $g_{\pm}$. %, $\tilde{g}_\pm$.
Instead,
\begin{equation}
(k^2 - \omega^2) \tilde{g}_\pm(\omega, k) \mp \frac{k \dot\theta_0 }{2} \left( e^{i t_0 m_a} \, \tilde{g}_\pm(\omega + m_a, k) + e^{-i t_0 m_a} \, \tilde{g}_\pm(\omega - m_a, k) \right) = 1
\label{eq:recursive}
\end{equation}
couples $\tilde{g}_\pm(\omega)$ to $\tilde{g}_\pm(\omega \pm m_a)$.
This complication stems from the fact that the solutions of the homogeneous equations of motion are Mathieu functions.

In both of the limits $\dot\theta_0 T \gg 1$ and $\dot\theta_0 T \ll 1$, the leading forms of the Green functions can be extracted from \eqref{eq:recursive} without invoking Mathieu functions or their Fourier transforms.
Section~\ref{sec:resonant} focuses on the former limit, in which frequencies $\omega = \frac{1}{2} m_a \pm \mathcal O(\dot\theta_0)$ are resonantly enhanced, inducing exponential growth for these unstable frequencies.
In the alternate limit $\dot\theta_0 T  \ll 1$ where the exponential growth is not realized, and for $\dot\theta_0 T \gtrsim 1$ for signals that do not include the resonantly enhanced frequencies, Section~\ref{sec:notresonant} provides a continued fraction expression for the Green function that converges quickly for small $\dot\theta_0 \ll m_a$.

\subsection{Resonantly Enhanced Propagation} \label{sec:resonant}

For small $\dot\theta_0 \ll m_a$, \eqref{eq:recursive} suggests that the solution for $\tilde{g}_\pm(\omega)$ could be found as an expansion in $k\dot\theta_0/(k^2 - \omega^2)$. However, for $\omega \approx k \pm\mathcal O(\dot\theta_0)$ the simple perturbative expansion is disrupted, especially near frequencies $\omega = \pm \frac{1}{2} m_a$ where $\omega^2 = (\omega \pm m_a)^2$.

At late times, when $\dot\theta_0 T \gtrsim 1$, the exponential growth of the unstable modes with $\omega = \pm m_a/2 + \mathcal O(\dot\theta_0)$ dominates the propagation of a signal.
In this late-time limit the Green function can be approximated by integrating over the resonantly enhanced modes,\footnote{Here we employ a multi-scale technique used in Ref.~\cite{VANEL201615} in an asymptotic analysis of lattice Green functions.}
\begin{align}
g_\pm(t | t_0, z) &\approx \sum_{k, \omega \approx \pm m_a/2} \int\! \frac{dk d\omega\, e^{i k z - i \omega (t - t_0) } }{(2\pi)^2} G(\omega_\pm, k_\pm) ,
\label{eq:Gwkappx}
\end{align}
where 
we expand $\omega$ and $k$ about $\pm m_a/2$, defining
\begin{align}
\epsilon \equiv \pm \frac{\dot\theta_0}{4 m_a},
&&
k_\pm \equiv \pm \frac{m_a}{2} + \alpha \epsilon,
&&
\omega_\pm \equiv \pm \frac{m_a}{2} + \beta \epsilon.
%&&
%k^2 - \omega^2 = \epsilon  (\alpha - \beta) \left( m_a + \epsilon \alpha + \alpha \beta \right).
\label{eq:epsilondef}
\end{align}
The factor of $\pm \dot\theta_0$ in \eqref{eq:greenDEosc} corresponding to right- and left-polarized light is absorbed into the definition of the parameter $\epsilon$.
%Near the resonance, the rescaled momenta $\alpha$ and $\beta$ take $\mathcal O( m_a)$ values,  and the function $\tilde{g}_\pm$ is $\mathcal O(\epsilon^{-1})$.
Even in the corner of ALP parameter space with small masses and relatively large couplings, \eqref{eq:dottheta} indicates that $\epsilon \lesssim 10^{-1}$ remains perturbatively small. For ALPs more closely resembling a QCD axion with $m_a > 10^{-12}\, \eV$,  $g_{a\gamma\gamma} < 10^{-11}\, \GeV^{-1}$, and fixed $\rho_a \approx 0.4\, \GeV/\text{cm}^3$, the value of $\epsilon$ drops to $\epsilon < 10^{-11}$. 

The function $G(\omega_\pm, k_\pm)$ introduced in \eqref{eq:Gwkappx} is defined to include only the $\mathcal O(\epsilon^{-1})$ part of $\tilde{g}_\pm$ in the neighborhood of the resonant frequencies, where $\alpha$ and $\beta$ are $\mathcal O(m_a)$.
By dropping the $\mathcal O(\epsilon^{0})$ portion of the Green function, \eqref{eq:recursive} can be disentangled to solve for $G(\omega_\pm, k_\pm)$, %with the result
\begin{align}
G\left(\omega_\pm, k_+ \right) 
= \frac{1}{\epsilon m_a} \left( \frac{ \alpha \pm \beta + m_a e^{\mp i t_0 m_a} }{\alpha^2 - \beta^2 - m_a^2 } \right)
&&
G\left(\omega_\pm, k_- \right) 
= -\frac{1}{\epsilon m_a} \left( \frac{ \alpha \mp \beta + m_a e^{\mp i t_0 m_a} }{\alpha^2 - \beta^2 - m_a^2 } \right).
\end{align}
%For $\alpha^2 < m_a^2$, the imaginary poles induce exponential growth at late times $T = t- t_0$ when $\epsilon m_a T \gg 1$.
%Defining
%\begin{align}
%\beta_0 \equiv + \sqrt{\alpha^2 - m_a^2} = + i \sqrt{m_a^2 - \alpha^2}
%\end{align}
In terms of $\alpha$ and $\beta$, \eqref{eq:Gwkappx} reduces to
\begin{align}
 g_\pm(t|t_0,z)  
\approx & \int\!\frac{4\epsilon^2 d\alpha d\beta\, e^{i \epsilon \alpha z - i \epsilon \beta(t- t_0)} }{(2\pi)^2 \epsilon m_a (\alpha^2 - \beta^2 - m_a ^2)}
 \times
\Bigg( i \alpha \cos \frac{m_a(t - t_0)}{2} \sin \frac{m_a z}{2} - i \beta \sin \frac{m_a(t-t_0)}{2} \cos\frac{m_a z}{2} 
\nonumber\\ & ~~~~~~~~~~ + i m_a  \cos \frac{m_a (t+t_0)}{2}  \sin\frac{m_a z}{2} 
\Bigg) + \mathcal O(\epsilon^0). 
\label{eq:gpmlateappx}
\end{align}
As always, the retarded Green function is defined to satisfy $g(t < t_0) = 0$, so the contour in $\omega(\beta)$ passes above both poles at $\beta = \pm \sqrt{\alpha^2 - m_a^2}$, even when $\alpha^2 < m_a^2$.

Fortuitously, this integral is nearly identical to the one encountered in Section~\ref{sec:static}.
Following the example of Eqs.~(\ref{eq:magic}--\ref{eq:lambda}), we introduce the coordinate transformations
\begin{align}
\alpha \equiv m_a \cosh \varphi ,
&&
\beta_0 \equiv + \sqrt{\alpha^2 - m_a^2} = + i \sqrt{m_a^2 - \alpha^2 }  \equiv m_a \sinh \varphi,
\end{align}
\begin{align}
%\pm m_a \epsilon z = \sqrt{\lambda} \sinh \eta
m_a \epsilon z = \sqrt{\lambda} \sinh \eta ,
&&
m_a \epsilon (t - t_0) = \sqrt{\lambda} \cosh\eta ,
&&
\lambda = m_a^2 \epsilon^2 ( (t - t_0)^2 -  z^2).
\end{align}
As in Figure~\ref{fig:contour}, $\varphi$ is such that $\cosh\varphi$ runs smoothly from $-\infty$ to $+\infty$ with $\text{Im}\cosh(\varphi) = 0$.
Extending the limits of integration in \eqref{eq:gpmlateappx} to $-\infty< \beta < \infty$ and $-\infty< \alpha < \infty$, we find
\begin{align}
g(t|t_0, z) &\approx \frac{\Theta(t-t_0) \Theta((t-t_0)^2 - z^2)}{2} \Bigg(1 +  4 i \epsilon \cos\frac{m_a(t+t_0)}{2} \sin \frac{m_a z}{2} I_0(\sqrt{\lambda}) 
\nonumber\\&+ 4 \epsilon \left[ m_a \epsilon (t-t_0) \sin\frac{m_a(t-t_0)}{2} \cos \frac{m_az}{2} - m_a \epsilon z \cos\frac{m_a(t-t_0)}{2} \sin \frac{m_a z}{2} \right] \frac{I_1(\sqrt{\lambda}) }{\sqrt{\lambda}} \Bigg),
\label{eq:corrected}
\end{align}
which is again causal and exhibits propagation inside the lightcone. 
The $\epsilon^0$ term in the expansion is calculated separately, by considering the Maxwell theory $\epsilon \rightarrow 0$ limit. 
In the late-time limit, 
when $T = (t - t_0)$ satisfies
\begin{equation}
2 \epsilon m_a T + \log \epsilon \gg \log m_a T,
\label{eq:epsiloncondition}
\end{equation}
the $\epsilon I_0(\sqrt{\lambda})$ and $\epsilon I_1(\sqrt{\lambda})$ contributions become larger than $\mathcal O(1)$, and the resonantly enhanced modes dominate the Green function. 
%From the asymptotic form of the Bessel functions~\eqref{eq:besselasymptotic}, this occurs at times $T = (t - t_0)$ when
%\begin{equation}
%2 \epsilon m_a T + \log \epsilon \gg \log m_a T.
%\label{eq:epsiloncondition}
%\end{equation}
%When this condition is satisfied, the $\mathcal O(\epsilon^0)$ component of the Green function is small compared to the contribution from the resonant modes.

\medskip

Compared to the $m_a \rightarrow 0$ limit from Section~\ref{sec:static}, the approximate retarded Green function in the resonantly enhanced regime of $m_a \gg \dot\theta$ is remarkably similar.
The exponential growth scale $\mu(t)$ has been replaced by 
\begin{equation}
m_a \epsilon = \frac{\mu_0}{2} = \frac{\dot\theta_0}{4},
\end{equation}
which is in line with what we naively expect from Section~\ref{sec:static}. 
For example, if we used the steady-state result to approximate the late-time exponential growth by replacing $|\mu(t)|$ with its average value,
 $\langle | \mu(t) | \rangle = \frac{1}{\pi} \dot\theta_0$, 
the resulting estimate for the growth factor is off by only $27\%$.

Recall from \eqref{eq:dottheta} that for fixed axion density $\rho_a = 0.4\, \GeV/\text{cm}^3$,
  even in the corner of parameter space saturating $g_{a\gamma\gamma} \lesssim 10^{-10}\, \GeV^{-1}$ and $m_a \gtrsim 2 \cdot 10^{-21}\, \eV$, the value of $\epsilon$ is still perturbatively small,  $|\epsilon| \lesssim 2\cdot 10^{-3}$.
For this roughly-maximal value of $\epsilon$, \eqref{eq:epsiloncondition} is satisfied by $m_a T \gtrsim 3600$.
Elsewhere in the $(m_a, g_{a\gamma\gamma})$ parameter space, $\epsilon$ can assume significantly smaller values, requiring larger $m_a T \gg 10^3$ to satisfy \eqref{eq:epsiloncondition}.

Our treatment of the axion background as a \emph{coherently} oscillating field requires the field to remain coherent throughout the signal propagation, or $T < T_c$ with $T_c \sim (m_a v^2)^{-1}$, where $v \sim 10^{-3}$ is the virial velocity of the axions. 
In the case of photons traveling freely through space, as opposed to reflecting within some cavity, the propagation distance $L$ must also be smaller than some $L_c \sim (m_a v)^{-1}$.
For $\epsilon \lesssim 10^{-5}$, 
the onset of exponential growth indicated by \eqref{eq:epsiloncondition} requires $m_a T > 10^6 \sim 1/v^2$, meaning that decoherence effects become important on the timescales associated with the exponential growth, and must be accounted for. 
This result is consistent with Ref.~\cite{Arza:2020eik}, which found that for axion models with $10^{-8}\, \eV < m_a $, decoherence completely obscures the exponential growth.
Additionally, for these models the width is narrow enough that gravitational redshift by the dark matter halo is sufficient to detune the resonance, even if the  velocity for the axion cloud is taken to be $v \ll 10^{-3}$ in order to satisfy $m_a T > 1/v^2$,
leading the authors of Ref.~\cite{Arza:2020eik} to conclude that for $m_a > 10^{-8}\, \eV$ in the observable range the parametric resonance at $\omega  = \frac{1}{2} m_a$ never develops into exponential growth.

Nevertheless, there is a window where $\epsilon \gtrsim 10^{-5}$ occupying a couple decades of the $(m_a, g_{a\gamma\gamma})$ parameter space in which the resonance can develop. It applies to extremely low-frequency radiation of $\omega \gtrsim 10^{-21}\, \eV$, or equivalently
$f \gtrsim  10^{-7}\, \text{Hz} \sim 10\, \text{yr}^{-1}$, which does not propagate through the interstellar medium.
%\bl{The prospects of creating and observing radiation at such extremely low frequencies are considered in Section~\ref{sec:pheno}, e.g. orbiting charged BH?}

%At extremely late times, where $\epsilon^2 m_a T \gtrsim O(1)$, the contributions from the narrow resonances at $\omega = \pm m_a + \mathcal O(\epsilon^2 m_a)$ become relevant as well.
%For the phenomenologically viable values of $\epsilon$ and $m_a T$, this 

\subsection{Propagation Without Resonance} \label{sec:notresonant}

In the opposite limit to \eqref{eq:epsiloncondition}, where $\epsilon m_a T\sim \dot\theta T \ll 1$ and the resonance is not given time to grow, an alternate approach provides a perturbative expansion of the Green function in powers of $\epsilon$.
This approach is also valid in the $\epsilon m_a T > 1$ limit for signals that do not include support within the instability band $\omega = m_a/2 \pm \mathcal O(\dot\theta)$, and generically for any signal where the resonant component from \eqref{eq:corrected} can be approximated by its series expansion.

For values of $\omega \neq m_a/2$, \eqref{eq:recursive} can be rearranged into a continued fraction solution for $\tilde{g}_\pm(\omega)$, as a generalization of the relation
\begin{equation}
 \tilde{g}_\pm(\omega, k)  = \frac{1}{k^2 - \omega^2} + \frac{2 k m_a \epsilon }{k^2 - \omega^2} \left( e^{i t_0 m_a} \, \tilde{g}_\pm(\omega + m_a, k) + e^{-i t_0 m_a} \, \tilde{g}_\pm(\omega - m_a, k) \right) ,
\end{equation}
in terms of $\epsilon$ from \eqref{eq:epsilondef}.
By iterating the replacement of $\tilde{g}_\pm(\omega \pm m_a)$ with $\tilde{g}_\pm(\omega)$ and $\tilde{g}_\pm(\omega \pm 2 m_a)$,
we derive an approximation of the form $\tilde{g}_\pm(\omega, k) \approx a_\ell(\omega, k)$ as follows:
\begin{align}
\tilde{g}_\pm(\omega) &= a_0(\omega ) + b_0(\omega) \tilde{g}_\pm(\omega + m_a) + b_0(\omega)\tilde{g}_\pm(\omega - m_a) , \nonumber\\
\tilde{g}_\pm(\omega) &= a_1(\omega) + b_1(\omega) \tilde{g}_\pm(\omega + 2 m_a) + b_{-1}(\omega)\tilde{g}_\pm(\omega - 2 m_a) ,\nonumber\\
%\tilde{g}_\pm(\omega) &= a_2(\omega) + b_2(\omega) \tilde{g}_\pm(\omega + 4 m_a) + b_{-2}(\omega)\tilde{g}_\pm(\omega - 4 m_a) \\
		&~ \vdots \label{eq:tryna}\\
\tilde{g}_\pm(\omega) &= a_\ell(\omega) + b_\ell(\omega) \tilde{g}_\pm(\omega + 2^\ell m_a) + b_{-\ell}(\omega)\tilde{g}_\pm(\omega - 2^\ell m_a) ,
\nonumber
\end{align}
where the $k$ dependence of each function $\tilde{g}_\pm(\omega \pm 2^\ell m_a, k)$ has been left implicit, and where
\begin{align}
a_0(\omega) &= \frac{1 }{k^2 - \omega^2} ,
&
a_1(\omega) &= \frac{a_0(\omega) +  b_{+0}(\omega) a_0(\omega + m_a) + b_{-0}(\omega) a_0(\omega - m_a)  }{1 -  b_{+0}(\omega) b_{-0}(\omega + m_a ) + b_{-0}(\omega) b_{+0}(\omega - m_a)  } ,
\nonumber\\
b_{\pm0}(\omega) &= \frac{2\epsilon k m_a e^{\pm i m_a t_0}}{k^2 - \omega^2 } ,
&
b_{\pm 1}(\omega) & = \frac{b_{\pm0}(\omega) b_{\pm0}(\omega \pm m_a) }{1 -  b_{+0}(\omega) b_{-0}(\omega + m_a ) + b_{-0}(\omega) b_{+0}(\omega - m_a)  } .
\label{eq:beenaught}
\end{align}
The recursion is provided by
\begin{align}
a_{\ell+1}(\omega) &= \frac{a_\ell(\omega) + b_{+\ell}(\omega) a_\ell(\omega + 2^\ell m_a) + b_{-\ell}(\omega) a_\ell(\omega - 2^\ell m_a)  }{1 - b_{+\ell}(\omega) b_{-\ell}(\omega + 2^\ell m_a ) - b_{-\ell}(\omega) b_{+\ell}(\omega - 2^\ell m_a )  }, 
\nonumber
\\
b_{\pm(\ell+1)}(\omega) &= \frac{ b_{\pm \ell}(\omega) b_{\pm \ell} (\omega \pm 2^\ell m_a)  }{1 - b_{+\ell}(\omega) b_{-\ell}(\omega + 2^\ell m_a ) - b_{-\ell}(\omega) b_{+\ell}(\omega - 2^\ell m_a )  } .
\label{eq:continuedA}
\end{align}

The convergence of the continued fraction expression effectively depends on a small-$b$ expansion, 
and for generic values of $\omega \sim k \gg O(\epsilon m_a)$ counting powers of $\epsilon$ is relatively easy: all of the $a_i$ are $\mathcal O(\epsilon^0)$, while $b_{\pm \ell} \sim \mathcal O(\epsilon^{2^\ell})$.
Near the $k\approx \omega$ poles, where $k = \pm \omega + \mathcal O(\epsilon m_a)$, a factor of $(k^2 - \omega^2) \propto \epsilon^{-1}$ modifies the power counting to $b_{\pm \ell} \sim \mathcal O(\epsilon^{2^\ell -1})$.

However, when $\omega \approx \frac{n}{2} m_a$ for integer $n$, the power counting is overturned near the $k^2 = \omega^2$ poles, prompting the special treatment in Section~\ref{sec:resonant}.
For example, for $\omega = \frac{1}{2} m_a \pm \mathcal O(\epsilon m_a)$, the poles in $a_0(\omega)$ and $a_0(\omega - m_a)$ are encountered simultaneously, 
and rather than finding $a_1(\omega ) = a_0(\omega) (1 + \mathcal O(\epsilon) ) $, the difference between $a_1(\omega)$ and $a_0(\omega)$ becomes $\mathcal O(\epsilon^0)$; similarly, $b_{-1}(\omega \approx m_a/2) \sim \mathcal O(\epsilon^0)$.
Excepting this $\omega = \pm \frac{1}{2} m_a$ resonance and the family of higher-order, narrower resonances, 
the Green function can otherwise be approximated to arbitrary order in $\epsilon$ by
\begin{equation}
\tilde{g}_\pm(\omega, k) = a_\ell(\omega, k)  + \mathcal O(\epsilon^{2^\ell - 1}).
\end{equation}

The $\ell = 0$ case is simply the Maxwell theory $\epsilon \rightarrow 0$ result. At $\ell = 2$, the expression for the retarded Green function at $\mathcal O(\epsilon^2)$ is
%\begin{align}
%g_\pm(t|t_0, z) &= \left( \frac{\Theta(z) \Theta( t- t_0 - z) }{2} +\frac{\Theta(-z) \Theta( t- t_0 + z) }{2}  \right) \left[ 1 \pm i \theta_0 \cos\frac{m_a(t + t_0) }{2}\sin\frac{m_a z}{2} + \mathcal O(\theta_0^2) \right] ,
%%\left[ 1 \mp i \theta_0 \sin\frac{m_a(t + t_0) }{2}\cos\frac{m_a z}{2} + \mathcal O(\theta_0^2) \right],
%\end{align}
\begin{align}
g_\pm(t|t_0,z) &= \left( \frac{1}{2}\Theta(z) \Theta( t - t_0 - z) + \frac{1}{2}\Theta(-z) \Theta( t - t_0 + z)\right) \Bigg( 1 
%\nonumber\\&
+ \epsilon \, 4 i \cos\left(\frac{m_a (t + t_0) }{2}\right) \sin \left(\frac{m_a z}{2}\right) 
\nonumber\\&
+ \epsilon^2 \bigg[ -4 + 4 \cos( m_a( t + t_0) ) \cos(m_a z) + 2 m_a (t - t_0) \cos\left(\frac{m_a z}{2}\right) \sin \left(\frac{m_a( t -  t_0) }{2} \right)
\nonumber\\&
-2 \cos \left(\frac{ m_a (t - t_0) }{2}\right) \left( \left[ -2 + 2 \cos\left( m_a ( t + t_0) \right) \right]\cos \left(\frac{m_a z}{2}\right) + m_a z \sin \left(\frac{m_a z}{2} \right) \right)
\bigg] 
\nonumber\\&
+ \mathcal O(\epsilon^3) \Bigg),
\label{eq:continued2}
\end{align}
%where retarded Green function is defined by the contour in $\omega$ that lies above all of the poles, and where the $k$ contour lies above the $k = - \omega$ and $k = -\omega \pm m_a$ poles, and below the $k = \omega$ and $k = \omega \pm m_a$ poles in the complex plane. In this way the Green function vanishes for $t < t_0$, and waves with positive (negative) phase and group velocities propagate in the positive (negative) $z$ directions.
which agrees at $\mathcal O(\epsilon)$ and even at $\mathcal O(\epsilon^2 m_a T)$ and $\mathcal O(\epsilon^2 m_a z)$ with the late-time expression \eqref{eq:corrected}.
Note that the series expansions of the Bessel $I_\nu(z)$ functions are
\begin{align}
I_0(\sqrt{z}) = 1 + \frac{z}{4} + \frac{z^2}{64} + \frac{z^3}{2304} + \ldots ,
&&
\frac{2}{\sqrt{z}} I_1(\sqrt{z}) = 1 + \frac{z}{8} + \frac{z^2}{192} + \frac{z^3}{9216} + \ldots.
\label{eq:besselseries}
\end{align}

To express $g_\pm$ in terms of $\dot\theta_0$, recall that a $\pm$ sign is incorporated into the definition of $\epsilon$, where
\begin{align}
g_+(t|t_0, z): \epsilon \rightarrow + \frac{\dot\theta_0}{4},
&&
g_-(t|t_0, z): \epsilon \rightarrow - \frac{\dot\theta_0}{4}.
\end{align}
As a result, the replacement $\epsilon \rightarrow -\epsilon$ is equivalent to switching $g_+ \leftrightarrow g_-$; taking the complex conjugate, $g_\pm^\star = g_\mp$; or applying the parity transformation $z \rightarrow -z$.

In Appendix~\ref{sec:cubic} we provide the $\mathcal O(\epsilon^3)$ form of $g(t | t_0, z)$. 
We also demonstrate that as $ m_a T \gg 1$ and $ m_a z \gg 1$ approach the late-time limit, the leading $\epsilon (\epsilon m_a T)^n$ and $\epsilon (\epsilon m_a z)^n$ terms reconstruct the series expansions of $I_0(\sqrt{\lambda})$ and $I_1(\sqrt{\lambda})/\sqrt{\lambda}$.
In \eqref{eq:trustandverify} we verify this explicitly as far as the $1/192$ coefficient of the $\epsilon (\epsilon m_a T)^5$ term.
This indicates that the continued fraction expression for the Green function provides a smooth interpolation between the $\epsilon m_a T \ll1$ and $\epsilon m_a T \gg 1$ limits. % as long as $\epsilon < 1$.

\medskip

By including the $\mathcal O(\epsilon^2)$ terms, \eqref{eq:continued2} is more precise than \eqref{eq:corrected} in the $\epsilon m_a t \ll 1$ limit, 
and it includes a novel effect: the $- 4 \epsilon^2$ term in the expansion, which is not proportional to any sinusoidal factors.
When the Green function is convolved with a signal of some duration $T$ and some spectrum of frequencies $\Omega$, in the $T \gg m_a^{-1}$ limit the sinusoidal terms act as approximate Dirac $\delta$ functions to enhance the modes with $\Omega \approx \frac{n}{2} m_a$ for integers $n \geq 1$.

If the axion mass is heavy enough that $m_a$ coincides with observable frequencies of light, then this resonant enhancement may be the most easily visible effect.
However, for very light ALP dark matter where $m_a\sim \omega < 2\pi \cdot \mathcal O(\text{kHz})$ corresponds to difficult-to-detect radio waves,   
the frequency-independent perturbation to the Green function becomes much more significant.
As $\epsilon^2$ is proportional to $\rho_a$, changes in the axion density can modify the strength of visible light passing through it, causing ``nongravitational microlensing.'' 

Depending on the cosmological history,  some fraction of the axions can clump together to form minihalos with density perturbations $\delta \rho_a / \rho_a$ potentially much larger than $\mathcal O(1)$~\cite{Kolb:1994fi}. (For recent work, see~\cite{Nelson:2018via,Visinelli:2018wza,Buschmann:2019icd,Eggemeier:2019khm,Blinov:2019jqc}.)
In addition to the gravitational microlensing, the direct effect from axion electrodynamics on starlight passing through an axion cluster may be detectable if the average value of $\epsilon^2 = \frac{1}{16} g_{a\gamma\gamma}^2 \rho_a/m_a^2$ is not too small.

\subsection{Numeric Results} \label{sec:oscnumeric}

%\bl{For this section: compare numeric solution to DE, and e.g. $a_{\ell = 2, 3}(\omega)$ solution from non-resonant expansion?}
\begin{figure}
\centering
\def\wid{0.49}
\includegraphics[width=\wid\textwidth]{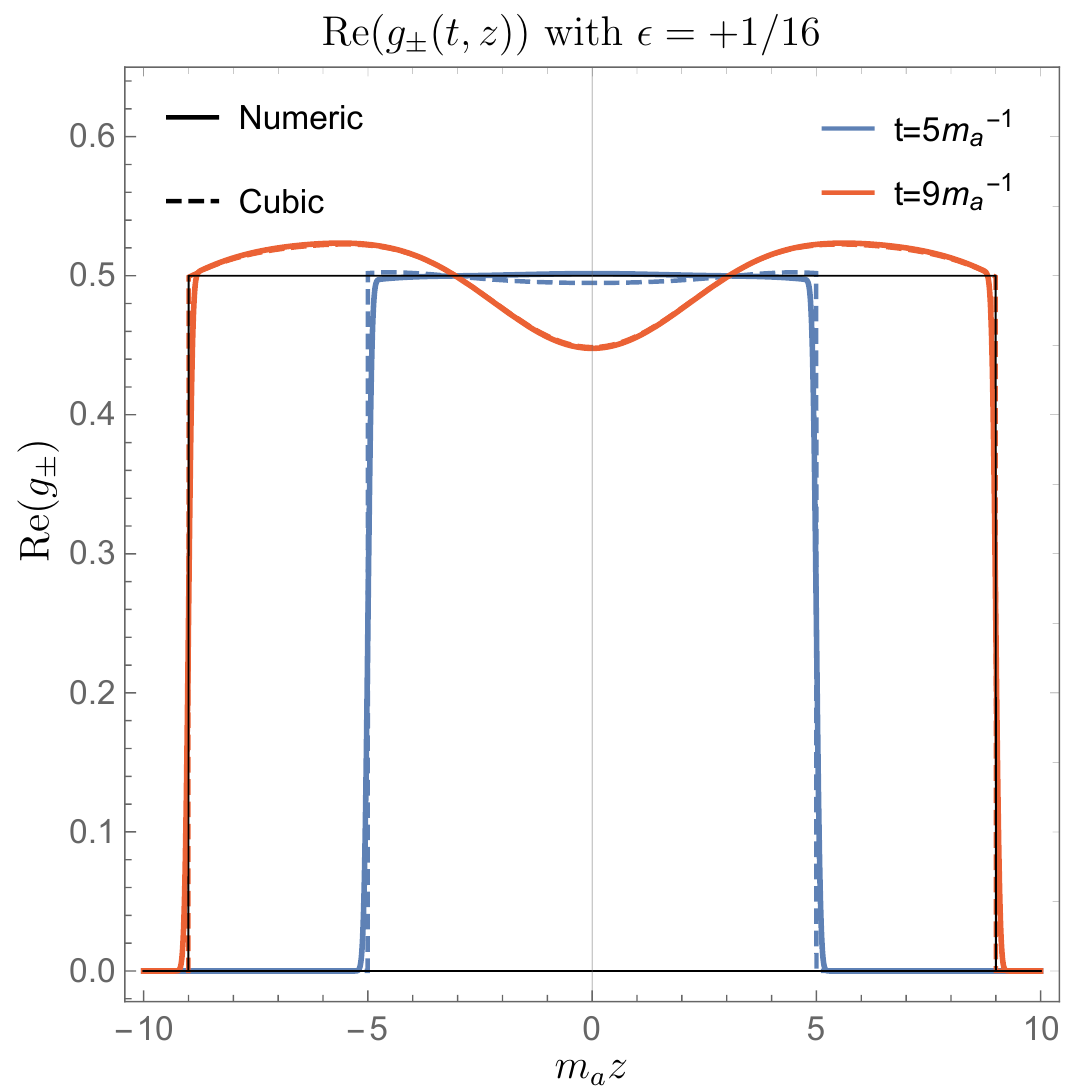}
\includegraphics[width=\wid\textwidth]{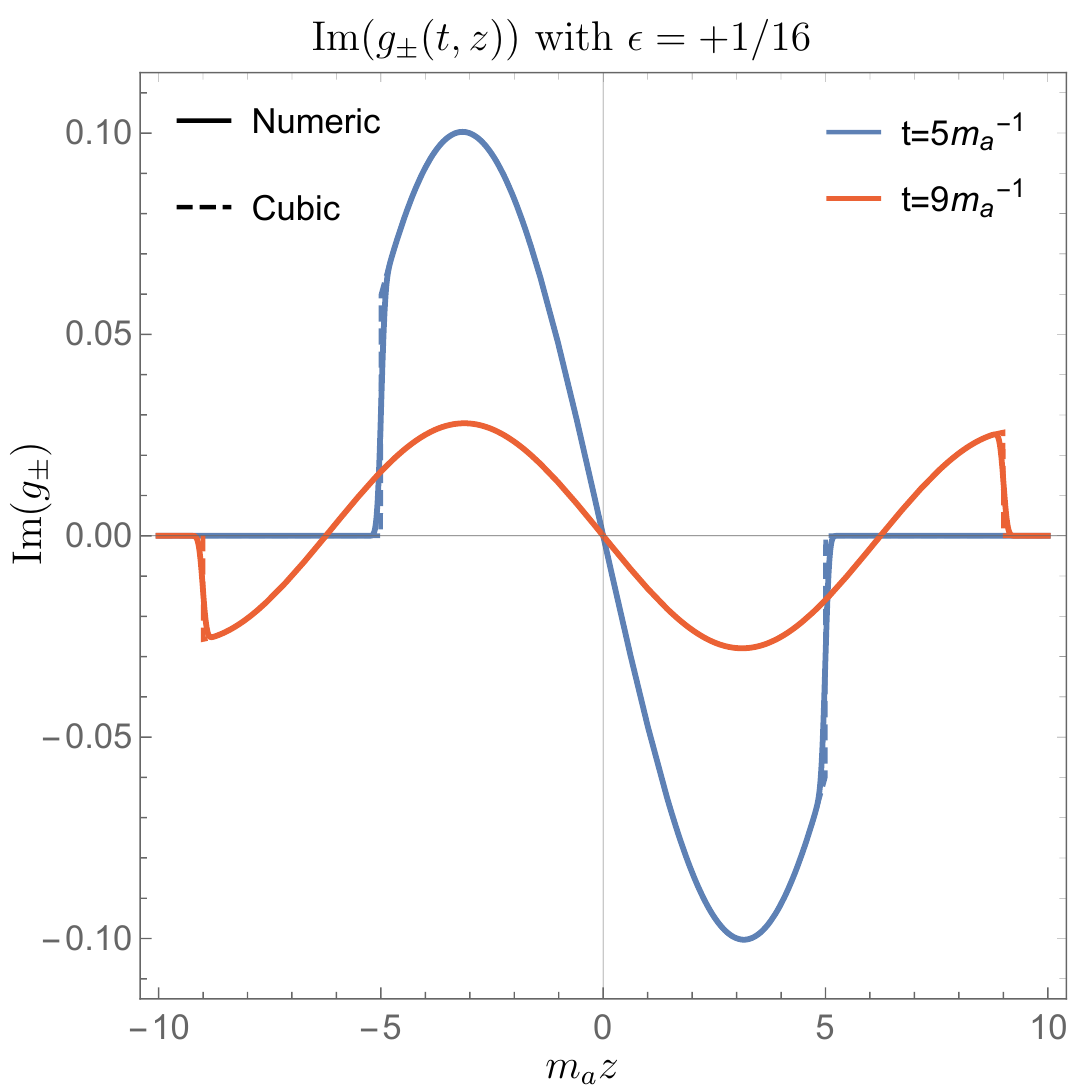}

\includegraphics[width=\wid\textwidth]{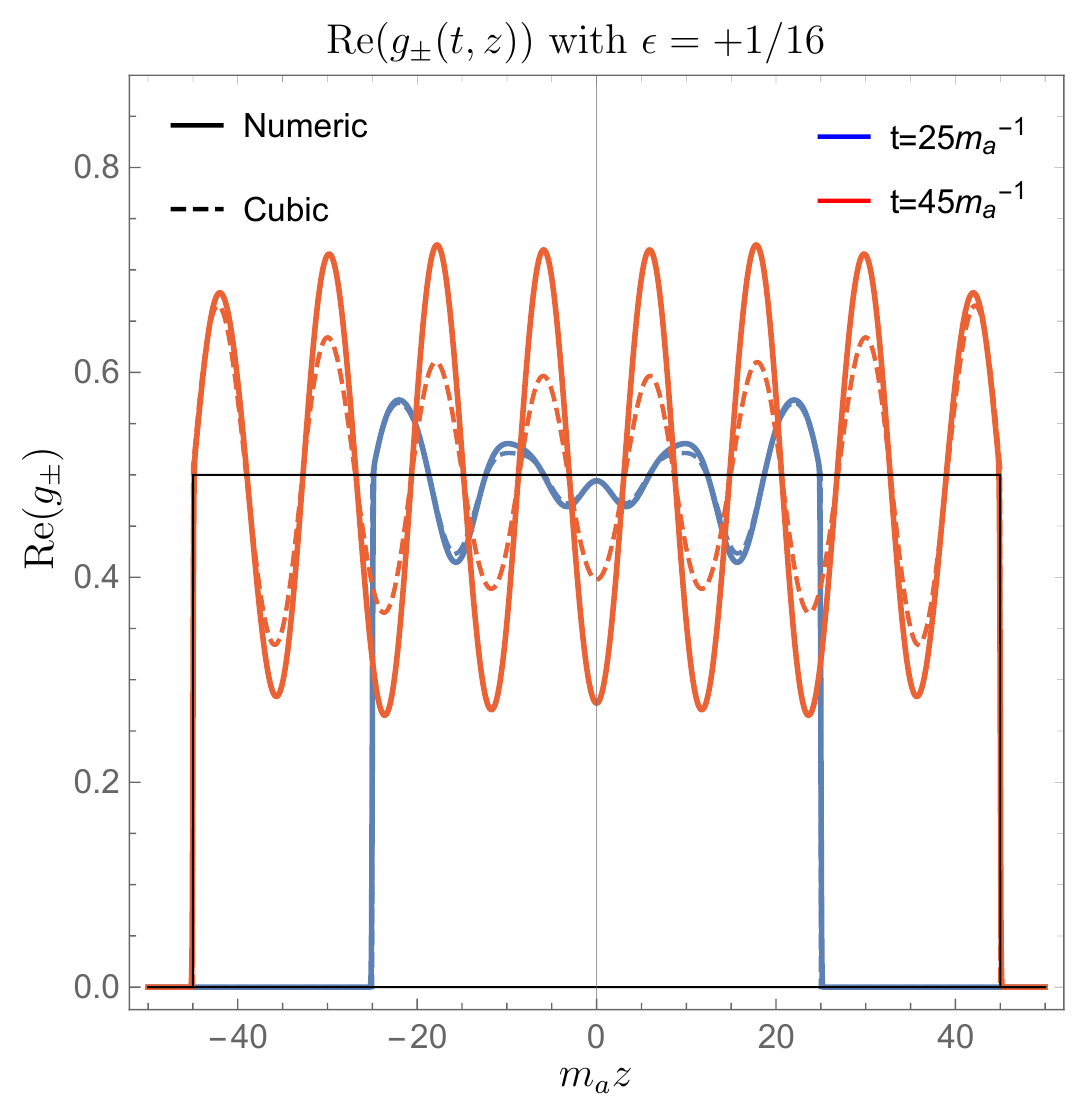}
\includegraphics[width=\wid\textwidth]{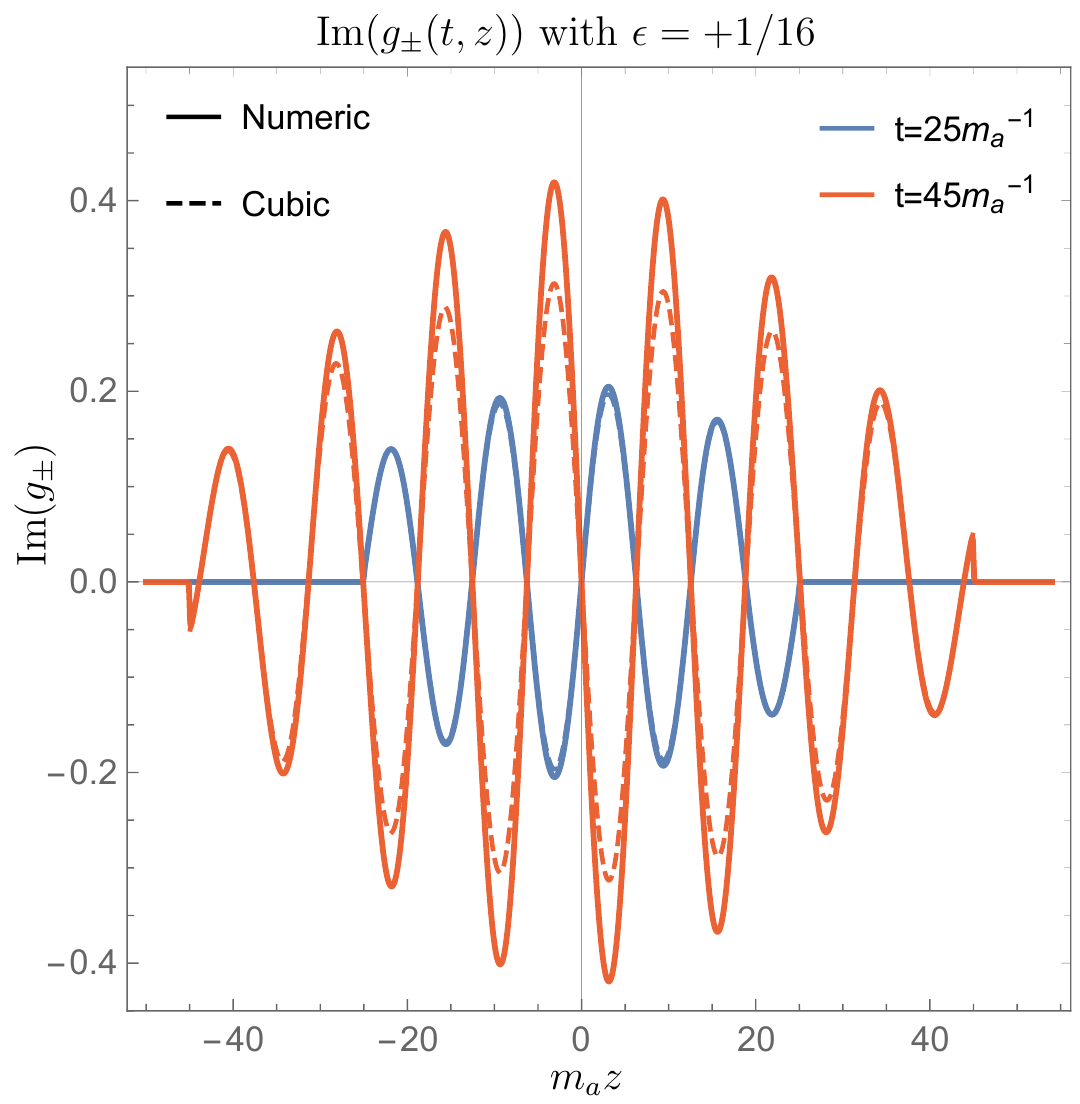}
\caption{Left column: The real part of the Green function $\text{Re}(g_\pm(t|t_0, z))$ is plotted as a function of $z$, comparing the series expansion \eqref{eq:cubicappx} at $\mathcal O(\epsilon^3)$ (dashed) against the numeric result (solid) at the following moments in time: $m_a t = \{5,9\}$ (top), $m_a t = \{ 25,  45\}$ (bottom).
A thin black line shows the $\epsilon\equiv \pm \frac{1}{4 }\dot\theta_0/m_a \rightarrow 0$ result at $t = 9 m_a^{-1}$ and $t = 45 m_a^{-1}$, in the upper and lower panels (respectively).
Right column: the series expansion (dashed) and numeric result (solid) for $\text{Im} (g_\pm(t|t_0, z))$ are shown as functions of $z$ for the same fixed values of $t$ as in the left column, at early times $t = \{ 5 m_a^{-1}, 9 m_a^{-1} \}$ (top) and late times $t = \{ 25 m_a^{-1} , 45 m_a^{-1} \}$ (bottom). 
In the $\epsilon \rightarrow 0$ limit (ordinary Maxwell theory), $\text{Im}(g_\pm) \rightarrow 0$.
Changing the sign of $\epsilon$ is equivalent to replacing $g_\pm$ with its complex conjugate, $g_\pm \rightarrow g_\pm^\star$, or performing the parity transformation $z \rightarrow - z$.
In this example we take $\epsilon = + 1/16$ and $t_0 = 0$, so that the top and bottom rows represent ``early'' and ``late'' times, with $\epsilon m_a t \ll 1$ and $\epsilon m_a t \gtrsim 1$, respectively. 
The $t = 45 m_a^{-1}$ curve in red shows the breakdown of the series expansion for $\epsilon m_a t \gg 1$.
%A small finite value of $\sigma = 0.05 m_a^{-1}$ was used for the numeric computation.
}
\label{fig:cubic}
\end{figure}

In Sections~\ref{sec:resonant} and~\ref{sec:notresonant} we used  different methods to approximate the Green function in the homogeneous, oscillating axion background. In this section we verify \eqref{eq:corrected} and \eqref{eq:continued2} by comparing them to the numeric solution of the differential equation
\begin{equation}
\Big(\partial_t^2 - \partial_z^2 + 4 i m_a \epsilon  \cos(m_a t) \, \partial_z \Big)  g_\pm(t - t_0, z - z_0) = \lim_{\sigma \rightarrow 0} 
\frac{1 }{2\pi \sigma^2} \exp \left(\frac{-(t-t_0)^2}{2 \sigma^2} \right) \exp \left(\frac{-(z-z_0)^2}{2 \sigma^2} \right).
\label{eq:greenDEnumeric}
\end{equation} 
(Recall that $\epsilon \equiv \pm \frac{\dot\theta_0}{4 m_a}$.) For the numeric calculations, we use a small value of $\sigma < \epsilon m_a^{-1}$ to approximate the delta function source.

\begin{figure}[h]
\centering
\includegraphics[width=0.49\textwidth]{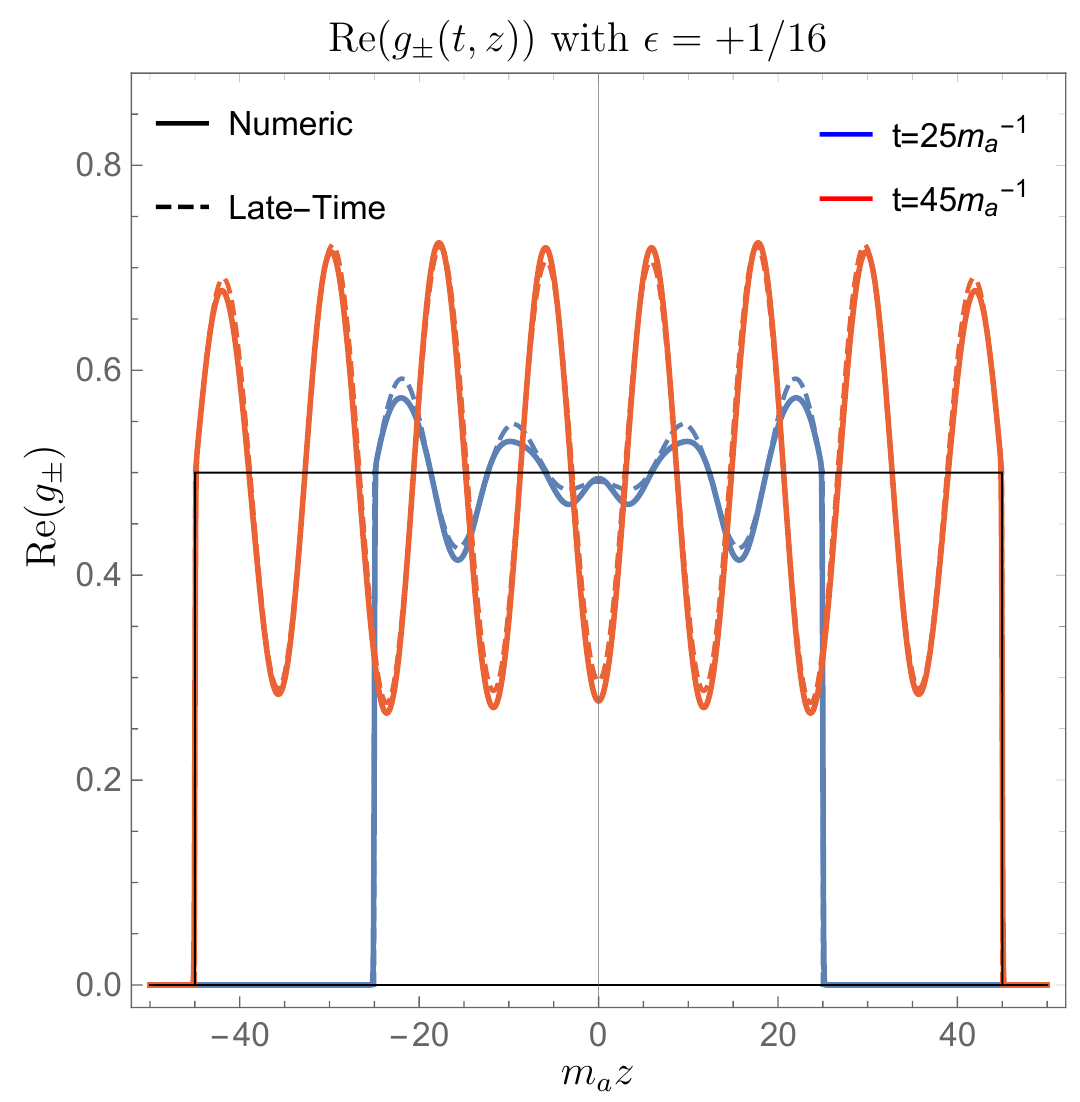}
\includegraphics[width=0.49\textwidth]{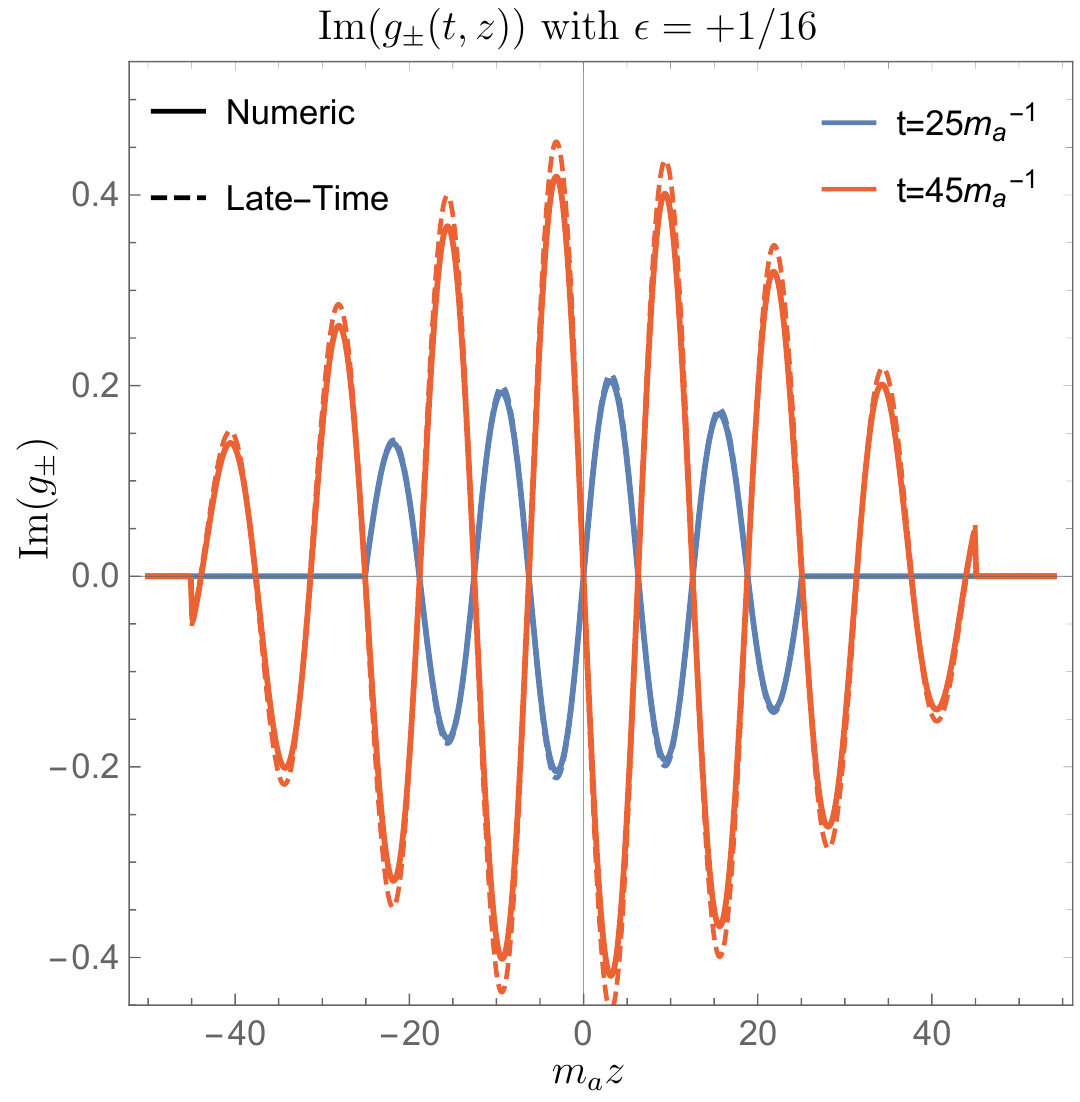}
\caption{The real (left) and imaginary (right) parts of the Green function $g(t|t_0, z)$ are plotted as functions of $z$ at fixed values of $m_a t = \{ 25,  45\}$. In each plot the late-time approximation \eqref{eq:corrected} (dashed) closely matches the numeric solution (solid), improving on the non-resonant series expansions used in the lower panels of the previous figure.
In this example, $\epsilon \equiv \pm \frac{\dot\theta_0}{4 m_a} = + 1/16$ and $t_0 = 0$, and the numeric calculation was performed with $\sigma = 0.05 m_a^{-1}$.
%Due to the oscillations in $t$, the magnitude of $\text{Im}(g_\pm)$ is greater at $t = 25 m_a^{-1}$ than it is at $t = 35 m_a^{-1}$.
For reference, the $\epsilon = 0$ solution for $\text{Re}(g_\pm)$ is shown at $t = 45 m_a^{-1}$ (thin black line).
}
\label{fig:res-late}
\end{figure}

In Figure~\ref{fig:cubic}, we compare the series expansion of the continued fraction $\tilde{g}_\pm (\omega, k) = a_{\ell }(\omega, k)$ at $\ell = 2$ to the numeric result calculated from \eqref{eq:greenDEnumeric} with $\sigma = 0.05 m_a^{-1}$. 
Rather than truncating the series expansion at $\mathcal O(\epsilon^2)$ as in \eqref{eq:continued2}, we use the $\mathcal O(\epsilon^3)$ expression \eqref{eq:cubicappx} from Appendix~\ref{sec:cubic}.
For simplicity, we use $t_0 = 0$ in this example, with $\epsilon = 1/16$.
The agreement at early times, $\epsilon m_a t \ll 1$, is quite good, but the approximation begins to fail by $t \gtrsim 2 m_a^{-1}/\epsilon$. 
Even though the $\mathcal O(\epsilon^3)$ expression includes terms that grow as $\epsilon (\epsilon m_a T)$ and $i \epsilon (\epsilon m_a T)^2$, by $t \geq 45 m_a^{-1}$ the exponential growth has begun to invalidate the series expansion in $(\epsilon m_a T)$ for both the real and imaginary parts of the Green function. 
%{\bf\color{blue}PD: Why is the red dashed line so close to the solid line in the upper panels? Accident? (blue and purple look like they're getting worse, but then red is good again?) Also, in the lower panels, particularly on the left, things are looking a little overcrowded. Maybe just use the red and blue times? I don't feel strongly. Same for fig 6.}

Figure~\ref{fig:res-late} shows the superior agreement between the late-time expression \eqref{eq:corrected} and the numeric result at later times $\epsilon m_a t \sim \mathcal O(1)$.
Compared to the continued fraction solution \eqref{eq:cubicappx}, \eqref{eq:corrected} is missing the $\mathcal O(\epsilon^2 (\epsilon m_a T)^0)$ and $\mathcal O(\epsilon^3 (\epsilon m_a T)^1)$ terms, making it less precise at early times. 
However, because \eqref{eq:corrected} includes terms of $\mathcal O(\epsilon (\epsilon m_a T)^n)$  for all nonnegative integers $n$, it is the correct choice for handling $\mathcal O(1)$ values of $\epsilon m_a T$.

\medskip

At extremely late times, when $m_a T \gg \epsilon^{-2}$, even our ``late-time'' approximation from Section~\ref{sec:resonant} fails to capture the dominant behavior of the Green function.
In addition to the $\omega = \pm m_a/2$ poles, the contribution from the next-to-leading resonance at $\omega = \pm m_a + \mathcal O(\epsilon^2 m_a)$ becomes significant.
To calculate the extremely-late Green function, the methods of Section~\ref{sec:resonant} can be repeated with some new $k_\pm = \pm m_a + \epsilon^2 \gamma$ and $\omega_\pm = \pm m_a + \epsilon^2 \delta$. 
Together with the subleading terms from the $\omega = \pm m_a/2$ poles, the resulting expression would include all terms of $\mathcal O(\epsilon^2 (m_a T)^n)$, and should match the $n=0$ and $n=1$ terms that appear already in \eqref{eq:cubicappx}.
For the phenomenologically viable values of $\epsilon$ with $\rho_a \sim 0.4\, \GeV/\text{cm}^3$, the decoherence at $T_c \sim (m_a v^2)^{-1}$ ensures that $\epsilon^2 m_a T \lesssim \mathcal O(1)$ for all $T \lesssim T_c$, and the axion field loses coherence before the behavior at these ``extremely late'' times can be explored.

\section{Conclusions} \label{sec:conclusion}
In this paper we have computed a set of Green functions appropriate to various limits of axion electrodynamics. 
Between Section~\ref{sec:green2d}, Section~\ref{sec:resonant}, and Section~\ref{sec:notresonant}, 
our analysis covers the phenomenologically viable parameter space for coherent axion backgrounds, as well as more extreme ALP models with stronger couplings or enhanced energy densities.

Our study of photon propagation at early times $T \ll m_a^{-1}$ in Section~\ref{sec:static} shows that the QCD axion induces gentle modifications to the propagation of local disturbances in standard electrodynamics.
ALP models in the more extreme corners of parameter space can instigate more dramatic growth in low-frequency modes as signal pulses pass through space.
In Section~\ref{sec:osc}, extending our analysis to account for the oscillation of the axion field, we derive a Green function that preferentially enhances radiation with frequencies close to $\omega = n \frac{m_a}{2}$ for positive integers $n $.
%At late times these frequency modes experience exponential growth, beginning
In all cases, perhaps the sharpest qualitative distinction between radiation in Maxwell theory and in axion electrodynamics is the presence of inside-the-lightcone propagation in the latter.

We collect our main results below. For plane waves symmetric in the $x$ and $y$ directions, the Green functions $g_\pm$ satisfy the equations of motion
$(\partial_t^2 - \partial_z^2 \pm i \dot\theta(t) \partial_z)  g_\pm(t - t_0, z - z_0) = \delta(z - z_0) \delta(t - t_0)$
for right- and left-polarized light, respectively.

\medskip
%\begin{itemize} \item \textbf{Semi-static Limit}
%\paragraph{Semi-static Limit} 
$\bullet$ \textbf{Semi-static Limit}
 at times $t \ll m_a^{-1}$, with $\mu = \frac{1}{2} \dot\theta_0 \cos(m_a t) \simeq \textit{const}$:
\begin{align}
g_\pm(t-t_0, z) = e^{\pm i \mu z} \frac{\Theta(t-t_0) \Theta((t-t_0)^2 - z^2) }{2}  I_0 \left( \sqrt{\mu^2 ((t-t_0)^2 -  z^2)} \right).
\tag{\ref{eq:greensoln2d}}
\end{align}
The phase factor $e^{\pm i \mu z}$ provides polarization-dependent phase velocities, as anticipated by the dispersion relation \eqref{eq:phasev}. The exponentially growing $I_0$ function corresponds to superluminal group velocities for both polarizations of light.
For nonstandard ALP models with $m_a \ll \dot\theta_0$, the exponential growth at $ t \gtrsim \mu^{-1}$ can begin while the semi-static condition $t \ll m_a^{-1}$ is still satisfied.
Otherwise, when $\mu t \lesssim 1$, the Bessel function is well approximated by its series expansion in powers of $\mu t$ and $\mu z$.

\medskip
%\item \textbf{Resonant, Late-Time Limit} 
%\paragraph{Resonant, Late-Time Limit}
$\bullet$ \textbf{Resonant, Late-Time Limit}
for $|\epsilon| m_a t \gtrsim 1$, where $\epsilon m_a = \pm \frac{1}{4} \dot\theta_0$ for $g_\pm$, respectively:
\begin{align}
g_\pm(t|t_0, z) &= \frac{\Theta(t-t_0) \Theta((t-t_0)^2 - z^2)}{2} \Bigg(1 +  4 i \epsilon \,  I_0(\sqrt{\lambda})  \cos\left(\frac{m_a(t+t_0)}{2}\right) \sin\left( \frac{m_a z}{2}\right)
\nonumber
\\&~~+ 4 \epsilon  \frac{I_1(\sqrt{\lambda}) }{\sqrt{\lambda}} \bigg[ m_a \epsilon (t-t_0) \sin\left(\frac{m_a(t-t_0)}{2}\right) \cos\left( \frac{m_a z}{2}\right) 
\nonumber\\&~~
- m_a \epsilon z \cos\left(\frac{m_a(t-t_0)}{2} \right)\sin\left( \frac{m_a z}{2} \right) \bigg] 
%\nonumber\\&
%+ \mathcal O(\epsilon^2 (\epsilon m_a t)^n )
\Bigg),
\tag{\ref{eq:corrected}}
\end{align}
where $g_\pm$ is expressed as a function of $\lambda = \epsilon^2 m_a^2 ( (t-t_0)^2 - z^2)$.
This Green function describes the behavior at late times, $\lambda \gtrsim 1$, when the resonant enhancement of frequencies $\omega \simeq m_a/2$ is the dominant effect. 
In contrast to \eqref{eq:greensoln2d}, \eqref{eq:corrected} is appropriate for the standard ALP models with $\dot\theta_0 \ll m_a$, and is valid even for $t \gg m_a^{-1}$. 
Both  \eqref{eq:greensoln2d} and \eqref{eq:corrected}
% limits of the Green function 
exhibit exponential growth at late times $t \gg \dot\theta^{-1}$, though in the latter case the enhancement is specific to frequency modes within the narrow resonance $\omega = m_a/2 \pm \mathcal O(\dot\theta)$.

At extremely late times, when $\epsilon^2 m_a t \gg 1$, the contributions from the higher resonances such as $\omega = m_a \pm \mathcal O(\dot\theta^2/m_a)$ also become significant. 
Given the small values of $\epsilon$ for most regions of ALP parameter space, and the fact that the axion background usually exhibits decoherence before $\epsilon^2 m_a t \sim \mathcal O(1)$ is satisfied, we do not provide the Green function in this limit. 
However, the methods of Section~\ref{sec:resonant} can be extended in a straightforward manner to describe the dominant behavior in the $\epsilon^2 m_a t \gtrsim 1$ limit.

%\newpage
\medskip
%\item \textbf{Non-Resonant Propagation}
%\paragraph{Non-Resonant Propagation}
$\bullet$ \textbf{Non-Resonant Propagation}
 for $|\epsilon| m_a t \lesssim 1$, and for signals that do not include the resonant frequencies $\omega = m_a/2 + \mathcal O(\dot\theta_0)$: 
\begin{align}
g_\pm(t|t_0,z) &= \frac{\Theta(t-t_0) \Theta((t-t_0)^2 - z^2)}{2} \Bigg( 1 
%\nonumber\\&
+ \epsilon \, 4 i \cos\left(\frac{m_a (t + t_0) }{2}\right) \sin \left(\frac{m_a z}{2}\right) - 4 \epsilon^2
\nonumber\\&
+ \epsilon^2 \bigg[4 \cos( m_a( t + t_0) ) \cos(m_a z) + 2 m_a (t - t_0) \cos\left(\frac{m_a z}{2}\right) \sin \left(\frac{m_a( t -  t_0) }{2} \right)
\nonumber\\&
-2 \cos \left(\frac{ m_a (t - t_0) }{2}\right) \left( \left[ -2 + 2 \cos\left( m_a ( t + t_0) \right) \right]\cos \left(\frac{m_a z}{2}\right) + m_a z \sin \left(\frac{m_a z}{2} \right) \right)
\bigg] 
\nonumber\\&
+ \mathcal O(\epsilon^3) \Bigg),
\tag{\ref{eq:continued2}}
\end{align}
%\end{itemize}
where the $\mathcal O(\epsilon^3)$ term of the non-resonant series expansion is provided in \eqref{eq:cubicappx} in Appendix~\ref{sec:cubic}.
As long as $|\epsilon| m_a t < 1$ and $\epsilon \lesssim 1$, the Green function can be calculated to arbitrary precision in $\mathcal O(\epsilon^n)$ using this approach.
For $|\epsilon| m_a t \gtrsim 1$, this series expansion interpolates smoothly onto the Bessel functions in \eqref{eq:corrected}, which include all terms of order $\epsilon (\epsilon m_a t)^n$ and $\epsilon (\epsilon m_a z)^n$ for $n = 0, 1, 2 \ldots$.

In addition to the enhancement of specific frequencies $\omega \approx \frac{n}{2} m_a$, \eqref{eq:continued2} and \eqref{eq:cubicappx} include terms that are not sinusoidal, and which modify the propagation of all frequencies of light.
For ALP models with small axion masses, where the resonant frequencies themselves are too small to detect, the frequency independent $(1 - 4 \epsilon^2)$ part of the Green function continues to affect visible wavelengths of light,
and may provide a new signal of ALP dark matter in regions where the axion density $\rho_a \propto \epsilon^2$ varies significantly.

%{\bf\color{blue} PD: reproduce main equations. Discuss main properties and limitations, folding in the comments below.}

%assembled a coalition of Green functions, inspired by axion electrodynamics but applicable to any physical system governed by the equations of motion \eqref{eq:greenDE} or \eqref{eq:greenDEosc}.

%The standard hierarchy $m_a \gg \dot\theta_0$ is well-described by the continued fraction expression \eqref{eq:continued2} and its late-time resonant limit \eqref{eq:corrected}, which extend the range of validity out to $ m_a T \ll \epsilon^{-2}$.
%If it is necessary to consider very late times, where $ m_a T \gtrsim \epsilon^{-n}$ for some $n \geq 2$, the treatment of Section~\ref{sec:resonant} can be extended in a fairly straightforward manner.

Both analyses in Section~\ref{sec:osc} are predicated upon $\epsilon = \pm \frac{1}{4 } \dot\theta_0/m_a$ being a small parameter, and so the $\epsilon > \mathcal O(1)$ case is generally the most difficult to address.
If $\epsilon > 1$ then the continued fraction expression for the Green function does not converge. 
In terms of the resonant analysis in Section~\ref{sec:resonant}, once $\epsilon^n m_a t \gtrsim 1$, the Green function receives contributions from an infinite set of poles at $\omega_n = \frac{n}{2} m_a$.
This unusual  $m_a \ll \dot\theta$ limit is handled by the treatment in Section~\ref{sec:green2d} for times $t \ll m_a^{-1}$, where the semi-static approximation is valid.
However, once $ t \gg m_a^{-1}$, %are both much larger than unity, 
%the semi-static approximation is invalid. %, and if $\epsilon \gtrsim 1$ the continued fraction expression for the Green function does not converge. 
%In this case 
the $\epsilon > 1$ Green function must be written in terms of Mathieu functions or calculated numerically.

%{\bf\color{blue} Discuss what would be interesting for future work. Effects on laser pulses? Including spatial gradient effects?}

\medskip

If axions make up some component of the dark matter, then  their presence may be discerned through their influence on propagating photons.
% their presence may be discerned by their influence on photons
%Not only are the Green functions derived in this work of mathematical interest, 
Sensitive terrestrial experiments designed to measure the inside-the-lightcone propagation or the polarization-dependent perturbations from axion electrodynamics might complement existing detection strategies for ALP dark matter.
Astrophysical observations may also be sensitive to the effects of axions. 
Depending on the axion mass, the resonant enhancement and frequency-independent modifications could provide additional opportunities to detect axion dark matter, especially if some fraction of the dark matter has collapsed into minihalos.

%There are several opportunities for future research.
The work presented here can be developed in several directions. 
To fully understand the propagation of light through a galaxy containing axions,  especially in the neighborhood of axion minihalos,
it is important to consider the effects from spatial gradients in the axion background. 
For light propagating through multiple coherent patches, over distances $L \gtrsim (m_a v)^{-1}$ or for times $T \gtrsim (m_a v^2)^{-1}$,  decoherence effects are similarly important.
%Together with the Green functions derived in this work, further study of  assessment of the capabilities of existing and planned experiments, there are several opportunities for future research.
%With the Green functions derived in this work, we can assess the capabilities of potential detection techniques for a wide range of ALP models, shining a new light in the continuing search for dark matter. 
Finally, Green function techniques may provide useful tools for exploring the sensitivity of terrestrial axion detection experiments. We hope to return to these issues in future work.

\section*{Acknowledgements}
We are grateful to Carlos Blanco, Nico Fernandez, and Matt Reece for helpful conversations, and to Yoni Kahn for collaboration in the early stages of the work.
The work of PD and BL is supported in part by the National Science Foundation Grant No.\ PHY-1719642, and the work of PA and BL is supported in part by the US Department of Energy Grant No.\ DE-SC0015655.
PA acknowledges the hospitality of the Kavli Institute for Theoretical Physics, which is supported in part by the National Science Foundation under Grant No.\ NSF-PHY-1748958.
BL thanks the Simons Center for Geometry and Physics, Stony Brook University at which some of the research for this paper was performed.

\appendix

\section{Continued Fraction Green Function At Higher Order} \label{sec:cubic}

In Section~\ref{sec:notresonant}, we listed the first few terms of the continued fraction \eqref{eq:continuedA}, truncating the series expansion of $a_2(\omega, k)$ at $\mathcal O(\epsilon^2)$. 
The number of terms in the expansion increases rapidly for higher powers of $\epsilon$. For example, the $\mathcal O(\epsilon^3)$ expression for the Green function is
\begin{align}
g(t|t_0, z ) &= %\left( \frac{1}{2}\Theta(z) \Theta( t - t_0 - z) + \frac{1}{2}\Theta(-z) \Theta( t - t_0 + z)\right)
\frac{\Theta( t- t_0) \Theta( (t - t_0)^2 - z^2) }{2} 
 \Bigg( 1 + \epsilon \, 4 i \cos\left(\frac{m_a (t +  t_0 )}{2} \right) \sin\left( \frac{m_a z}{2} \right)
\nonumber\\&
+ \epsilon^2 \bigg[ -4 + 4 \cos( m_a (t +  t_0)) \cos(m_a z) + 2 m_a (t - t_0) \cos\left(\frac{m_a z}{2}\right) \sin \left(\frac{m_a( t -  t_0) }{2} \right)
\nonumber\\&~~~~
-2 \cos \left(\frac{ m_a (t -  t_0 )}{2}\right) \left( \left[ -2 + 2 \cos\left( m_a (t +  t_0) \right) \right]\cos \left(\frac{m_a z}{2}\right) + m_a z \sin\left( \frac{m_a z}{2}\right) \right)
\bigg]
\nonumber\\&
+ \frac{i}{3} \epsilon^3 \bigg\{ \cos\left(\frac{m_a( t +  t_0 )}{2}\right) \bigg( 12 m_a z \cos \frac{m_a z}{2} \sin^2\left(\frac{m_a (t -  t_0 )}{2}\right) 
\nonumber\\&~~~~
+ \sin \left( \frac{m_a z}{2}\right) \Big[ 2 \cos(2 m_a t) - 38 \cos (m_a (t -  t_0) )
+ 2 \cos (2 m_a t_0) + 6 \cos (m_a (t +  t_0) )
\nonumber\\&~~~~
- 63 - 6 m_a (t - t_0) \sin (m_a( t -  t_0)) + 3 m_a^2 (t - t_0)^2 - 3 m_a^2 z^2 \Big]
\bigg)
\nonumber\\&~~~~
+8 \sin (m_a z) \bigg[ 3 \cos (m_a t) + 3 \cos (m_a t_0) - \cos(m_a (2  t +  t_0) )- \cos(m_a ( t + 2  t_0)) \bigg]
\nonumber\\&~~~~
+ 9 \cos\left(\frac{3m_a (t +   t_0) }{2} \right) \sin\left( \frac{3 m_a z}{2} \right)
\bigg\}
+ \mathcal O(\epsilon^4)
\Bigg).
\label{eq:cubicappx}
\end{align}
To calculate $g_\pm(t|t_0, z)$ to $\mathcal O(\epsilon^4)$--$\mathcal O(\epsilon^7)$, it becomes necessary to integrate the series expansion of $a_{\ell = 3}(\omega, k)$ rather than $a_2(\omega, k)$, which can become tedious.
 At $\mathcal O(\epsilon^4)$, for example, 
 $a_3(\omega, k)$ includes triple poles $(k^2 - \omega^2)^3$; double poles $(k^2 - (\omega \pm m_a)^2)^2$ and $(k^2 - (\omega \pm 2 m_a)^2)^2$;
and single poles $(k^2 - (\omega \pm 3m_a)^2)$ and $(k^2 - (\omega \pm 4 m_a)^2)$.
After integrating over $k$, the $\mathcal O(\epsilon^4)$ component yields quadruple poles at $\omega = \pm \frac{1}{2} m_a$, triple poles at $\omega = \pm \frac{3}{2} m_a$, double poles at $\omega = \pm m_a$ and $ \omega = \pm \frac{5}{2} m_a$, and single poles at $\omega = 0$, $\omega = \pm 2 m_a$, $\omega = \pm 3 m_a$, and $\omega = \pm \frac{7}{2} m_a$.

Already at $\mathcal O(\epsilon^3)$, we can see that the $\mathcal O(\epsilon (\epsilon m_a T)^n)$ terms for $n=0, 1,2$ match the series expansions of the Bessel functions $I_0$ and $I_1$ given in \eqref{eq:besselseries}, an important consistency check between Section~\ref{sec:resonant} and Section~\ref{sec:notresonant}.
By focusing exclusively on the poles at $k^2 = \omega^2$ and $k^2 = (\omega \pm m_a)^2$, we can verify \eqref{eq:corrected} by comparing it to the $\mathcal O(\epsilon (\epsilon m_a T)^n)$ expansion for arbitrarily large $n$. 
For example, at $n=5$, we integrate the $\mathcal O(\epsilon^6)$ expression for $a_3(\omega, k)$ to find that
\begin{align}
g(t|t_0, z) &= \left( \frac{1}{2}\Theta(z) \Theta( t - t_0 - z) + \frac{1}{2}\Theta(-z) \Theta( t - t_0 + z)\right) \Bigg( 
1 + 4 i \epsilon \cos\left(\frac{m_a (t +  t_0) }{2}\right) \sin \left(\frac{m_a z}{2} \right)
\nonumber\\
& 
\times \left( 1 + \frac{\epsilon^2}{4} \bigg( m_a^2 (t - t_0)^2 - m_a^2 z^2 \bigg) + \frac{\epsilon^4}{64} \bigg( m_a^2 (t - t_0)^2 - m_a^2 z^2 \bigg)^2 + \ldots \right)
\nonumber\\&
+ 2 \epsilon^2 \left(  m_a (t - t_0) \cos\left(\frac{m_a z}{2}\right) \sin\left( \frac{m_a (t -t_0) }{2}\right)  - m_a z \sin \left(\frac{m_a z}{2} \right) \cos \left(\frac{ m_a (t -t_0) }{2} \right)   \right) 
\nonumber\\& 
\times
\left(1 + \frac{\epsilon^2}{8} \bigg( m_a^2 (t - t_0)^2 - m_a^2 z^2 \bigg) + \frac{\epsilon^4}{192} \bigg( m_a^2 (t - t_0)^2 - m_a^2 z^2 \bigg)^2 + \ldots  \right)
\Bigg)
\nonumber\\& 
 + \mathcal O(\epsilon^2 (\epsilon m_a T)^n),
\label{eq:trustandverify}
\end{align}
in complete agreement with \eqref{eq:corrected}.

%For optimal precision, the late-time resonant expression \eqref{eq:corrected} can be combined with the remaining fixed-order $\mathcal O(\epsilon^n)$ 

\section{Integral for Four Dimensional Green Function} \label{appx:integral}

The part of the Green function $G_{ij}$ with the $k_i k_j/k^2$ tensor structure cannot be easily expressed in terms of hypergeometric functions.
The problematic integral can be expressed as
\begin{equation}
A_{ij} =  \frac{1}{4\pi} \partial_i \partial_j \left[ \frac{1}{2} \Theta(t) \Theta(t^2 - r^2) \frac{2}{\m r} \mathcal I(t, r)\right]
\end{equation}
where
\begin{equation}
\mathcal I = \frac{\m r}{2} \int_{-1}^1 \! dq\, \cos(q r \m) I_0\left(\sqrt{\m^2 t^2 - \m^2 q^2 r^2 } \right) = \int_0^{\m r} d(q \m r)\, \cos(q r \m)  \, I_0\left(\sqrt{(\m t)^2 - (\m q r)^2 } \right).
\end{equation}
Defining the dimensionless parameters
\begin{align}
%\tau = \m t && \rho = q \m r.
v \equiv \frac{1}{4} q^2 \m^2 r^2
&&
v_0 \equiv \frac{1}{4} \m^2 r^2
&&
\sigma \equiv \frac{1}{4} \m^2 t^2
\end{align}
and replacing the trigonometric and Bessel functions with their equivalent hypergeometric functions,
\begin{align}
\mathcal I &= \int_0^{v_0} \frac{d v}{\sqrt{v}}\ {_0 F_1}\!\left.\left( \begin{array}{c} - \\ \frac{1}{2} \end{array} \right| - v \right) {_0 F_1}\! \left. \left( \begin{array}{c} - \\ 1 \end{array}  \right| \sigma-v  \right) 
 = \int_0^{v_0} \frac{d v}{\sqrt{v}}\ {_0 F_1}\!\left.\left( \begin{array}{c} - \\ \frac{1}{2} \end{array} \right| - v \right) \sum_{k=0}^\infty \frac{(\sigma - v)^k }{(k!)^2} \\
&= \int_0^{v_0} \frac{d v}{\sqrt{v}}\ {_0 F_1}\!\left.\left( \begin{array}{c} - \\ \frac{1}{2} \end{array} \right| - v \right) \sum_{k=0}^\infty \sum_{j = 0}^k 
\frac{\sigma^{k-j} (-v)^j}{k! j! (k-j)!} ,
\end{align}
the integral over $v$ can be completed:
\begin{align}
 \int_0^{v_0}\! d v\, v^{j - \frac{1}{2} } {_0 F_1}\!\left.\left( \begin{array}{c} - \\ \frac{1}{2} \end{array} \right| - v \right)
=
\frac{v_0^{j+\frac{1}{2}} }{j+\frac{1}{2} }\, {_1 F_2}\! \left.\left( \begin{array}{c} j+ \frac{1}{2} \\ \frac{1}{2}, \frac{j}{2} + \frac{1}{4}  \end{array} \right| - v_0 \right),
\end{align}
with the result
\begin{align}
\mathcal I &= \sum_{k = 0}^{\infty} \sum_{j=0}^k \frac{\sigma^{k-j} v_0^{j + \frac{1}{2} } (-1)^j  }{k! j! (k - j )! (j + \frac{1}{2} ) } \, {_1 F_2}\! \left.\left( \begin{array}{c} j+1 \\ \frac{1}{2}, \frac{j}{2} + \frac{1}{4}  \end{array} \right| - v_0 \right).
\end{align}
One of the two infinite series can be replaced by a hypergeometric function, after replacing the index $k$ with $\ell \equiv k - j$, so that both $j$ and $\ell$ run from zero to infinity:
\begin{align}
\mathcal I &=  \sqrt{v_0} \sum_{j=0}^\infty \sum_{\ell=0}^\infty  \frac{\sigma^\ell}{\ell! \Gamma(j+\ell+1) } \frac{(- v_0)^j }{j! (j + \frac{1}{2} )}   \, {_1 F_2}\! \left.\left( \begin{array}{c} j+\frac{1}{2} \\ \frac{1}{2}, \frac{j}{2} + \frac{1}{4}  \end{array} \right| - v_0 \right) \\
%&=  \sqrt{v_0} \sum_{j=0}^\infty \frac{(-v_0)^j}{j! (j+\frac{1}{2}) \Gamma(j+1)} \, {_0 F_1} \! \left.\left( \begin{array}{c} - \\ j+1 \end{array} \right| \sigma \right)   {_1 F_2}\! \left.\left( \begin{array}{c} j+\frac{1}{2} \\ \frac{1}{2}, \frac{j}{2} + \frac{1}{4}  \end{array} \right| - v_0 \right) \\
%%\mathcal I &= \sqrt{v_0} \sum_{j=0}^\infty \frac{\left(- \frac{1}{4} \m^2 r^2 \right)^j}{(j!)^2 (j+\frac{1}{2})} \, {_0 F_1} \! \left.\left( \begin{array}{c} - \\ j+1 \end{array} \right| \frac{\m^2 t^2}{4} \right)   {_1 F_2}\! \left.\left( \begin{array}{c} j+\frac{1}{2} \\ \frac{1}{2}, \frac{j}{2} + \frac{1}{4}  \end{array} \right| - \frac{1}{4} \m^2 r^2 \right).
\mathcal I &= \sqrt{v_0} \sum_{j=0}^\infty \frac{\left(- \frac{1}{4} \m^2 r^2 \right)^j}{j! (j+\frac{1}{2})} 
 \left(\frac{2}{\m t}\right)^j I_j(\m t)
\ {_1 F_2}\! \left.\left( \begin{array}{c} j+\frac{1}{2} \\ \frac{1}{2}, \frac{j}{2} + \frac{1}{4}  \end{array} \right| - \frac{1}{4} \m^2 r^2 \right).
\end{align}
Even for relatively large values of $\m t$ and $\m r$, the series expression for $\mathcal I$ converges relatively quickly.
In the asymptotic $v_0 \gg 1$ limit, the $_1 F_2$ function is approximately
\begin{equation}
\lim_{v_0 \rightarrow \infty}  {_1 F_2}\! \left.\left( \begin{array}{c} j+\frac{1}{2} \\ \frac{1}{2}, \frac{j}{2} + \frac{1}{4}  \end{array} \right| - v_0 \right)
\approx v_0^{j/4 + 1/8} \frac{2^{-j + 1/2} \sqrt{\pi} }{ \Gamma(3/4 + j/2) } \cos\left( 2 \sqrt{v_0} + \frac{\pi( 1 + 2 j) }{8} \right),
\end{equation}
while to leading order the Bessel functions $I_j(2\sqrt{\sigma})$ approach
\begin{equation}
\lim_{\m t \rightarrow \infty} I_j(\m t) \approx \frac{e^{\m t} }{ \sqrt{2\pi \m t} } \left( 1 - \frac{(4j^2 - 1)}{8 \m t} + \mathcal O\left( (\m t)^{-2} \right) \right).
\label{eq:besselasymptotic}
\end{equation}
Consequently, the number of terms in $\sum_{j=0}^{j_\text{max}}$ required for convergence is driven primarily by the value of $\m t$: in the limit $\m t \gg 1$, the series converges quickly for $j> j_\text{max}$ once
\begin{equation}
\frac{3 j_\text{max} }{2} \log \frac{j_\text{max} }{e} + j_\text{max} \log \frac{\m t}{\sqrt{2}} - \frac{5 j_\text{max}}{2} \log \frac{\m r}{2}  \gg \m t.
\end{equation}

\bibliography{references-AED.bib}

\end{document}